\documentclass[twocolumn]{autart}
\journal{...}

\usepackage{amsmath,amssymb}
\usepackage{MnSymbol}
\usepackage{lineno}
\usepackage{soul,color}
\usepackage{verbatim}
\usepackage{algpseudocode}
\usepackage{algorithm}
\usepackage{algorithmicx}
\usepackage{physics}
\usepackage {optidef}
\usepackage{subcaption}
\usepackage{float}
\usepackage{bm}
\usepackage{hyperref}
\hypersetup{
    colorlinks=true,
    linkcolor=blue,
    filecolor=magenta,      
    urlcolor=cyan,
    pdftitle={Overleaf Example},
    pdfpagemode=FullScreen,
    }

\makeatletter
\@ifundefined{thm}{\newtheorem{thm}{Theorem}}{}
\@ifundefined{cor}{}{}
\@ifundefined{lem}{\newtheorem{lem}{Lemma}}{}
\@ifundefined{claim}{}{}
\@ifundefined{axiom}{}{}
\@ifundefined{conj}{}{}
\@ifundefined{fact}{}{}
\@ifundefined{hypo}{}{}
\@ifundefined{assum}{}{}
\@ifundefined{prop}{\newtheorem{prop}[thm]{Proposition}}{}
\@ifundefined{crit}{}{}
\@ifundefined{defn}{\newtheorem{defn}[thm]{Definition}}{}
\@ifundefined{exmp}{\newtheorem{exmp}[thm]{Example}}{}
\@ifundefined{rem}{\newtheorem{rem}[thm]{Remark}}{}
\@ifundefined{prob}{\newtheorem{prob}[thm]{Problem}}{}
\@ifundefined{prin}{}{}
\@ifundefined{proof}{\newtheorem{proof}{Proof}}{}
\makeatother

\usepackage{tikz}
\usetikzlibrary{shapes.geometric, arrows}

\tikzstyle{startstop} = [rectangle, rounded corners, 
minimum width=2cm, 
minimum height=1cm,
text centered, 
draw=black, 
thick,
text width=3.5cm
]

\tikzstyle{io} = [trapezium, 
trapezium stretches=true, 
trapezium left angle=70, 
trapezium right angle=110, 
minimum width=3cm, 
minimum height=1cm, text centered, 
draw=black, fill=blue!30, ultra thick]

\tikzstyle{process} = [rectangle, 
minimum width=3cm, 
minimum height=1cm, 
text centered, 
text width=3cm, 
draw=black, ultra thick]

\tikzstyle{decision} = [diamond, 
minimum width=3cm, 
minimum height=1cm, 
text centered, 
draw=black, ultra thick]
\tikzstyle{arrow} = [ultra thick,->,>=stealth]

\usepackage{caption}
\usepackage{graphicx}

\usepackage{natbib}
\usepackage{xcolor}
\usepackage[inline]{enumitem}
\usepackage{float}
\usepackage{subfloat}

\begin{document}
\begin{frontmatter}
\title{Topology Reconstruction of a Resistor Network with Limited Boundary Measurements: An Optimization Approach} 
\author[Paestum]{Shivanagouda Biradar}\ead{eez198372@ee.iitd.ac.in}    
\author[Paestum]{Deepak U Patil}\ead{deepakpatil@ee.iitd.ac.in}        

\address[Paestum]{Indian Institute of Technology Delhi, India}         
\begin{abstract}                          
We address the problem of reconstructing the topology and edge resistance values of an unknown CPPR network using limited resistance distance measurements. A multistage topology reconstruction method is developed, assuming prior knowledge of the number of boundary and interior nodes, the maximum and minimum edge conductance, and the Kirchhoff index. The process begins by constructing a maximal circular planar network with edges composed of resistors and switches, excluding interior nodes. A sparse difference of convex programming problem $\mathbf{\Pi}_1$ combined with a round-down algorithm is formulated to determine optimal switch positions, yielding an initial topology. This topology serves as the basis for a heuristic method to place interior nodes. The heuristic involves reformulating $\mathbf{\Pi}_1$ into a new optimization problem $\mathbf{\Pi}_2$ with relaxed edge weight constraints and a quadratic cost function. The resulting placement of interior nodes may produce a non-planar topology. To address this, we develop a modified Auslander, Parter, and Goldstein algorithm to generate a set of planar network topologies. For each planar topology, edge weights are re-optimized by solving a third optimization problem $\mathbf{\Pi}_3$. These optimization problems are formulated as difference of convex programming problems, incorporating triangle inequality and Kalmanson’s inequality constraints. A numerical example demonstrates the effectiveness of the proposed method.
\end{abstract}
\end{frontmatter}

\section{Introduction}\label{sec:introduction}
Electrical networks are ubiquitous in daily life. Mechanical systems \cite{akbaba2022electric}, biological systems \cite{gomez2012modeling}, water distribution system \cite{veldman2015towards}, geological system \cite{jirkuu2019resistor}, and many fields use electrical networks to model the system and simplify analysis. In particular, resistor networks hold an important place in modeling different physical systems, such as modeling fractures in crystalline rocks \cite{jirkuu2019resistor}, the electrical resistivity of carbon composites \cite{lundstrom2022resistor}, soft robotics sensor arrays \cite{zhao2023measuring}, modeling graphene sheets and carbon nanotubes \cite{cheianov2007random}, Mott spiking neurons \cite{rocco2022exponential}, fluid transport networks \cite{zhu2006network} \& phylogenetic networks \cite{forcey2020phylogenetic}. In most practical cases, the network structure is often unavailable for analysis.\\
\indent The primary objectives of electrical network topology reconstruction are: $i)$ determining the network structure, and $ii)$ estimating the edge conductances using available boundary measurements. Reconstructing the topology of resistor networks is challenging due to their static nature and the limited availability of boundary and interior measurements \cite{dorfler2012kron}. In \cite{curtis1994finding}, the authors address the reconstruction of circular resistor networks represented as $C(m,n)$, where $m$ denotes the number of concentric circles and $n$ represents the number of rays originating from $n$ boundary nodes on the outermost circle. This structure is assumed to be known, with all boundary terminals available for measurements. The edge resistance values are computed using the response matrix and boundary measurements. Similarly, in \cite{curtis1990determining}, an algorithm is presented to compute edge resistances for general rectangular networks, assuming the network structure and boundary node accessibility are known. A gamma harmonic function based on Kirchhoff's law is used to calculate the resistances. However, in many practical scenarios, the network structure is unknown, and not all boundary terminals are accessible. For instance, in soft resistive sensor arrays, no structural information is available apriori, and only limited boundary measurements can be collected. In such cases, reconstructing the network becomes significantly more challenging due to the lack of interior node information. In \cite{curtis2000inverse}, the authors solve the reconstruction problem for well-connected, critical, circular, and planar networks, assuming full boundary terminal accessibility. Their approach involves computing disjoint paths using non-negative circular minors of the response matrix, which are then used to construct a medial graph identifying interior node positions. While effective, this method relies heavily on the availability of the full response matrix and assumes the network is critical and well-connected; restrictions that limit its applicability in real-world cases where only partial boundary data is available. Similar challenges arise in phylogenetics, where genetic distances analogous to resistance distances are used to reconstruct phylogenetic networks \cite{forcey2020phylogenetic}. Here too, it is assumed that the response matrix or equivalent data is fully known. In \cite{ghosh2008minimizing}, a convex optimization-based method is proposed to allocate edge weights, but this also assumes prior knowledge of the network structure. 

The topology reconstruction problem is also prevalent in power systems, where the goal is to identify the admittance matrix of a distribution network using voltage and current measurements at select nodes. The admittance matrix provides complete information about the network structure, and the radial (tree) structure of distribution networks simplifies identification. For instance, \cite{moffat2019unsupervised} estimates the admittance matrix using least squares and complex recursive grouping algorithms, while \cite{soumalas2017data} employs similar techniques. Similarly, in interconnected dynamical systems, topology reconstruction is achieved using time-series input/output data and model structures, as explored in \cite{van2021topology, nabi2012network, materassi2012problem}. {In \cite{zhou2022structure} authors investigates parameter identifiability and estimation for descriptor systems, where system matrices depend on parameters through a linear fractional transformation (LFT). Necessary and sufficient conditions for identifiability are derived \cite{zhou2024identifiability,zhou2022global}. These results eliminate the full-rank assumptions typically required in prior studies. In \cite{zhou2024frequency} metrics for absolute and relative sloppiness are introduced to evaluate parameter sensitivity in the frequency domain . An algorithm is proposed to determine a minimal set of frequencies at which frequency responses uniquely identify parameters.} \\\indent In our previous work \cite{biradar2023topology}, we developed a reconstruction algorithm for general circular planar resistor networks without assuming any prior knowledge of the network structure. Using Gr\"{o}bner basis method \cite{cox1994ideals}, the algorithm constructs the set of all networks consistent with a given response matrix, assuming no information about interior nodes is available. Additionally, in \cite{biradar2022topology}, we addressed the reconstruction of resistive networks with $1\Omega$ edge resistances using partially available resistance distance measurements. We also characterized a set of networks that satisfy these limited boundary measurements.\\
\indent {The circular planar structure is particularly well-suited for modeling real-world systems that exhibit layered or circular patterns, including geological resistivity models, biological networks, transportation systems, soft robotics sensor arrays, and radial power distribution networks.} Therefore, in this paper, we address the reconstruction of a general unknown CPPR network. We assume that only some boundary nodes are accessible for measurements, while all interior nodes are inaccessible. {This assumption reflects real-world challenges in fields like biomedical and geophysical systems, where accessing certain nodes is often impossible.} Additionally, the number of boundary and interior nodes, the maximum and minimum edge conductance, and the Kirchhoff index are known apriori. Importantly, no simplifying assumptions about the underlying network structure are made. The topology reconstruction process is divided into four stages, as explained below:
\begin{enumerate}

   \item {\textbf{Stage 1- Network Initialization:}} To begin, no information about the network topology is available. To construct an initial network, we design a network composed of resistors and switches. This process involves two key steps: first, we construct a maximal planar graph using only the boundary nodes. Next, each edge in this graph is replaced with a network of resistors and switches. The resistance of each edge is determined by the position of its associated switch, which can be either ``on'' or ``off.''

The objective at this stage is to determine an optimal combination of switch positions such that the resulting network closely matches the given resistance distance measurements. This optimization problem is formulated as a sparse difference of convex programming problem, denoted as $\mathbf{\Pi}_1$. The problem incorporates a quadratic cost function and employs a rounding-down algorithm to induce sparsity in the solution. Solving $\mathbf{\Pi}_1$ yields an initial network, $\Gamma_{aux}$.

It is important to note that this stage does not account for interior nodes. The placement of interior nodes within the initial network $\Gamma_{aux}$ is addressed in Stage 2 of the reconstruction process.
 \item \textbf{{Stage 2- Placement of Interior Nodes:}} In this stage, we introduce a heuristic method for placing $n_i$ interior nodes within the initial network $\Gamma_{\text{aux}}$. The method involves solving an optimization problem $\mathbf{\Pi}_2$, which is designed to determine the appropriate locations for some of the $n_i$ interior nodes along the edges of the network. The remaining interior nodes are classified as \emph{dangling nodes}, meaning no edges are connected to them. The optimization problem $\mathbf{\Pi}_2$ is a reformulation of $\mathbf{\Pi}_1$, with the key difference being the relaxation of constraints on edge conductances.
\item \textbf{{Stage 3- Constructing Planar Networks:}} After appropriately placing the interior nodes in the initial network $\Gamma_{\text{aux}}$, the connections among the interior nodes, as well as those between the interior nodes and the boundary nodes, remain unknown. To account for all possible internal connections in the unknown network, we initially connect each interior node to every other node in the network. Let this network be denoted as $\hat{\Gamma}$. However, such extensive interconnections may result in a non-planar network $\hat{\Gamma}$, which is inconsistent with the assumption that the target network is planar. To address this issue, it becomes necessary to extract a set of planar networks from the non-planar network $\hat{\Gamma}$. In this stage, we introduce a modified version of the Auslander, Parter, and Goldstein algorithm, which systematically constructs a set of planar networks from the given non-planar network. This ensures that the resulting networks adhere to the planarity constraint while preserving the structural properties of the original network.
\item \textbf{{Stage 4- Edge Weight Assignment:}} Finally, the edge weights of the constructed planar networks are determined by solving an optimization problem, $\mathbf{\Pi}_3$, which is similar in structure to $\mathbf{\Pi}_2$. This optimization ensures that the assigned edge weights satisfy the available resistance distance measurements and the Kirchhoff's index. From the resulting set of planar networks, we select the one that most closely matches the given resistance distance measurements and the Kirchhoff's index. This selected network represents the reconstructed CPPR network. 
\end{enumerate}
\subsection{Contributions}
\begin{enumerate}
\item In contrast to other works \cite{forcey2020phylogenetic},\cite{curtis2000inverse},\cite{biradar2022topology},\cite{ghosh2008minimizing}, 
\begin{enumerate*}
\item we assume that the available measurements are limited,
\item we explicitly account for the presence of interior nodes in the circuit, which are inaccessible for experiments, adding complexity to the reconstruction process,
\item more importantly, we only assume that the network structure is circular and planar, without imposing additional simplifying assumptions about the underlying graph of the unknown network. This makes our method applicable to a broader class of circular planar resistive networks.
\end{enumerate*}
\item To reconstruct an unknown CPPR network, we formulate the problem as a difference of convex programming problem. The objective function incorporates a quadratic cost, while the constraints enforce the triangle inequality and Kalmanson inequality on the resistance distances. These constraints naturally induce a difference of convex structure in the optimization problem, enabling us to leverage efficient numerical solvers
\item In the proposed algorithm, we adopt a novel strategy to construct an initial network, as outlined in  \textbf{Stage-1} . Specifically, we demonstrate that selecting an appropriate combination of switch positions based on resistance distances and the Kirchhoff index can be formulated as a difference of convex programming problem. Additionally, we provide a systematic method to generate an initial guess for the optimization solver, ensuring robust convergence.

For the placement of interior nodes, we develop a heuristic method that classifies some interior nodes as dangling nodes (nodes with no incident edges) and others as non-dangling nodes . This classification is achieved by solving a similar difference of convex programming problem, ensuring consistency with the network's physical properties.
    
\item We propose a modified Auslander, Parter, and Goldstein's planarity testing algorithm \cite{hopcroft1974efficient} to generate a set of planar electrical networks from a non-planar electrical network.
\end{enumerate}
\subsection{Mathematical Notations}
Let $S_1=\{a_1, a_2, \ldots, a_s\}$ be row indices, $S_2=\{b_1, b_2, \ldots, b_s\}$ be column indices, and let $M \in \mathbf{R}^{n \times n}$ be any arbitrary matrix, $M\left(S_1;S_2\right)$ be a submatrix formed from the set of row indices $S_1$ and the set of column indices $S_2$. $|\cdot|$ is the cardinality of the set. $\mathbf{R}^+$ is the set of positive real numbers and $\mathbf{Z}^+_{\le n}$ is the set of positive natural numbers up to value $n$. $\odot$ represents element wise multiplication. $\textbf{1}$ and $\textbf{0}$ is a vector of ones and zeros of appropriate dimension. $\mathcal{S}_m$ is a set of symmetric matrix of order $m$. $r^d_{i,j}$ is the resistance distance between nodes $i$ and $j$ and $r(ij)$ is the edge resistance of edge $ij$ in a network.  
\section{Problem Formulation}
Consider a $CPPR$ electrical network $\Gamma = (\mathcal{G}, \gamma)$, where $\mathcal{G} = (\mathcal{V}, \mathcal{E})$ represents a finite, simple, and connected circular planar graph. The graph $\mathcal{G}$ is embedded in a disc $D$ on the plane, which is bounded by a circle $C$. The set $\mathcal{V}$ is the set of the nodes of the graph, while set $\mathcal{E} \subseteq \mathcal{V} \times \mathcal{V}$ is the set of the edges connecting these nodes. The nodes are divided into two categories: boundary nodes, which lie on the circle $C$, and interior nodes, which are located within the disc $D$. Consequently, the set of all nodes can be expressed as $\mathcal{V} = \mathcal{V_B} \cup \mathcal{V_I}$, where $\mathcal{V_B}$ represents the set of boundary nodes and $\mathcal{V_I}$ represents the set of interior nodes. The number of boundary nodes is denoted by $|\mathcal{V_B}| = n_b$, and the number of interior nodes is denoted by $|\mathcal{V_I}| = n_i$. \emph{Both $n_b$ and $n_i$ are assumed to be known}. The boundary nodes in $\mathcal{V_B}$ are labeled from $1$ to $n_b$ in a clockwise circular order around the circle $C$. Let $\mathcal{A}$ denote the set of boundary nodes available for voltage and current measurements. The set of all nodes that are not accessible for measurements is defined as $\mathcal{U} = \mathcal{V} \setminus \mathcal{A}$. Among these inaccessible nodes, $\mathcal{U_B} \subseteq \mathcal{U}$ represents the subset of boundary nodes that are unavailable for measurements. The conductivity function $\gamma: \mathcal{E} \to \mathbb{R}^+$ assigns a positive real number $\gamma(\sigma)$ to each edge $\sigma \in \mathcal{E}$, representing the conductance of the edge. The resistance of the edge is then given by $r(\sigma) = \gamma(\sigma)^{-1}$, $\forall \sigma \in \mathcal{E}$. Let $\gamma_{\text{max}} := \max\{\gamma(\sigma): \forall \sigma \in \mathcal{E}\}$ and $\gamma_{\text{min}} := \min\{\gamma(\sigma): \forall \sigma \in \mathcal{E}\}$ denote the maximum and minimum edge conductances, respectively. {Such bounds are typically estimated using domain knowledge, material properties, or calibration measurements.} The resistance distance, denoted by $r^d_{i,j}$, is defined as the effective resistance measured between any two nodes $i, j \in \mathcal{V}$. The resistance distance matrix $\mathbf{R}_{\Gamma} \in \mathbb{R}^{m \times m}$, where $m = n_b + n_i$, encapsulates these distances, with entries defined as, $\mathbf{R}_{\Gamma}(i,j) = \mathbf{R}_{\Gamma}(j,i) = r^d_{i,j}, \quad \text{and} \quad \mathbf{R}_{\Gamma}(i,i) = 0, \quad \forall i,j \in \mathcal{V}.$ A related quantity, the Kirchhoff index, is the sum of the effective resistances across all pairs of nodes in the network:
\begin{equation}\label{eqn:kirchhoffindex}
 K_{\Gamma} := \frac{1}{2} \mathbf{1}^T \mathbf{R}_{\Gamma} \mathbf{1} = \sum_{s < t} r^d_{s,t}.   
\end{equation}
{The Kirchhoff's index, a global graph property, can be reasonably approximated using available resistance distance measurements and domain-specific knowledge.} Since the network is passive and contains no internal sources, the resistance distance between any two available boundary nodes can be determined experimentally by applying a known voltage across them and measuring the induced current. Specifically, the resistance distance is computed as the ratio of the applied voltage to the measured current. For the nodes in $\mathcal{A}$, the resistance distances are measured and collected into the set $r^d = \{r^d_{s,t} = \phi_{st}/i_s : \forall s,t \in \mathcal{A}\}$. This provides a submatrix of the resistance distance matrix $\mathbf{R}_{\Gamma}$, denoted as $\mathbf{R}_{\Gamma}(\mathcal{A}; \mathcal{A})$, whose entries are derived from $r^d$. Having established this, we now explore the relationship between the resistance distance matrix $\mathbf{R}_{\Gamma}$ and the structural properties of the network, as encoded in the Laplacian matrix $\mathcal{L}$, in the following section.
\subsection{Laplacian and Resistance Distance Matrix}
The Laplacian matrix $\mathcal{L}$ corresponding to any {electrical network $\Gamma$} is a symmetric $n\times n$ matrix $\mathcal{L}\left({\mathcal{G}}\right)$, defined as follows:
\begin{equation}\label{lap_def}
\left[\mathcal{L}\left(\mathcal{G}\right)\right]_{ij}=\left[\mathcal{L}_{ij}\right]\left\{
    \begin{array}{ll}
      = -\gamma\left(ij\right), & \mbox{if $ij\in \mathcal{E}$},\\
      = \sum\limits_{j\in \mathcal{N}\left( i \right)}{{\gamma\left(ij\right)}}, & \mbox{if $i=j$},\\
      =0, & \mbox{otherwise}.
    \end{array}
  \right.
\end{equation}   
It is shown in \cite{ghosh2008minimizing} that the resistance distance is related to the Laplacian matrix as follows:
\begin{equation}
   r^d_{i,j}  =\left[\mathcal{L}{\left( {\mathcal{G}} \right)^ \dagger }\right]_{ii} + \left[\mathcal{L}{\left( {\mathcal{G}} \right)^ \dagger }\right]_{jj} - 2\left[\mathcal{L}{\left( {\mathcal{G}} \right)^ \dagger }\right]_{ij} 
   \label{eqn:RDdef}
\end{equation}
where, $\mathcal{L(G)}^\dagger=\left(\mathcal{L(G)} + \frac{1}{n} \textbf{J}\right)^{-1}-\frac{1}{n}\textbf{J}$, $\textbf{J}=\mathbf{1}\mathbf{1}^T$, and $\mathbf{1}$ is vector of ones.
Using equation \eqref{eqn:RDdef}, we express $\textbf{R}_\Gamma$ as, 
\begin{equation}
    \textbf{R}_\Gamma = \textbf{J}\mbox{diag}\left(\mathcal{L(G)}^\dagger\right)+\mbox{diag}\left(\mathcal{L(G)}^\dagger\right) \textbf{J} - 2\mathcal{L\left(G\right)}^\dagger.
\end{equation} Further, let $\textbf{X}=\left(\mathcal{L(G)} + \frac{1}{n} \textbf{J}\right)^{-1}$ and $\bar{\textbf{X}}=\mbox{diag}\left(\mathcal{L(G)}^\dagger\right).$
 Then \begin{equation}\label{eqn:resis_equa}
     \textbf{R}_\Gamma = \textbf{J}\bar{\textbf{X}} + \bar{\textbf{X}}\textbf{J} - 2\textbf{X}.
\end{equation}
We now formulate the problem of reconstructing the topology and edge weights of an unknown CPPR network:
\begin{prob}
  Given the Kirchhoff index $K_{\Gamma}$, the number of boundary nodes $n_b$, the number of interior nodes $n_i$, the maximum and minimum edge conductances $(\gamma_{\text{max}}$ and $\gamma_{\text{min}})$, and the submatrix $\mathbf{R}_{\Gamma}(\mathcal{A}; \mathcal{A})$ obtained from measurements, perform the following tasks:
  \begin{enumerate}
      \item Estimate the full resistance distance matrix $\mathbf{R}_{\Gamma}$ corresponding to the unknown network $\Gamma$.
      \item Reconstruct the network topology and compute the edge weights $\gamma(\sigma)$, $\forall \sigma \in \mathcal{E}$ using the estimated $\mathbf{R}_{\Gamma}$. 
  \end{enumerate}
\end{prob} 
To address this problem, we leverage the relationship between the resistance distance matrix and the Laplacian matrix to infer the topology. Additionally, we utilize several properties of resistance distances to formulate intermediate optimization problems that facilitates the reconstruction of $\mathbf{R}_{\Gamma}$ from $\mathbf{R}_{\Gamma}(\mathcal{A}; \mathcal{A})$, which are discussed in the following section.
\subsection{Triangle Inequality \& Kalmanson's Inequality}
{Our primary objective is to reconstruct an unknown CPPR network using limited resistance distance measurements. To address missing boundary resistance distances, we enforce two key constraints: the triangle inequality ($\Delta \succeq 0$) and Kalmanson’s inequality ($\mathcal{K} \succeq 0$). The triangle inequality ensures that the effective resistance between any two nodes does not exceed the sum of resistances along alternative paths, preventing unrealistic shortcuts or negative resistances. Kalmanson’s inequality encodes the geometric properties of planarity, guaranteeing that crossing paths exhibit higher resistance than non-crossing paths, as required in planar networks. Together, these constraints guide the optimization process(formulated in section-3, 4 and 6), narrowing the solution space to physically meaningful and geometrically coherent networks, ensuring the reconstructed network is both structurally accurate and consistent with electrical principles.}

In a $\operatorname{CPPR}$ network, resistance distance satisfy the triangle inequality\cite{choi2019resistance}, as stated in Theorem \ref{lem:triineq}.
\begin{thm}\label{lem:triineq}\cite{choi2019resistance} 
For any three distinct \textbf{boundary nodes} $i,j,k$ in CPPR $\Gamma$ such that $1\le i < j < k \le n_b$, the resistance distances $r_{i,j}^d$, $r_{j,k}^d$ and $r_{i,k}^d$ satisfies $r_{i,k}^d \le r_{i,j}^d+r_{j,k}^d.$
\end{thm}
To enforce the triangle inequality as constraints, 
we select node indices $i, j, k$ such that at least 
one node belongs to the set $\mathcal{U_B}$, while the 
remaining nodes are chosen from the set of available nodes 
$\mathcal{A}$. Using these indices, we define the set:$\Delta  = \left\{ 
	\left( {r_{i,j}^d + r_{j,k}^d} \right) - r_{i,k}^d: i,j,k\, \mbox{is chosen as explained above}
\right\}.$ All elements of this set must be non-negative, which we denote as $\Delta \succeq 0$. Another critical property of resistance distances in CPPR networks is \emph{the Kalmansons property}, as described in Theorem \ref{kalmanson_thm}:
\begin{thm} \label{kalmanson_thm}\cite{forcey2020phylogenetic} 
For any four \textbf{boundary nodes} $i,j,k,l$ of CPPR $\Gamma$, satisfying $1 \le i<j<k<l \le n_b$, the resistance distances $r^d_{i,j}, r^d_{k,l}, r^d_{i,k}, r^d_{j,l}, r^d_{j,k}, \mbox{ and }r^d_{i,l}$  satisfy ${r^d_{i,k}} + {r^d_{j,l}} \ge {r^d_{i,j}} + {r^d_{k,l}} \, and \,
{r^d_{i,k}} + {r^d_{j,l}} \ge {r^d_{j,k}} + {r^d_{i,l}}.$
\end{thm}
\noindent To enforce Kalmanson’s inequalities as constraints, we first select valid boundary node indices $i, j, k, l \in \{a, b, c, d : 1 \leq i < a < b < c < d \leq n_b\}$, ensuring that at least one index corresponds to a node in $\mathcal{U_B}$, with the remaining indices chosen from $\mathcal{A}$. The following inequalities are then imposed: $({r^d_{i,k}} + {r^d_{j,l}}) - \left( {{r^d_{i,j}} + {r^d_{k,l}}} \right) \ge 0,\,
({r^d_{i,k}} + {r^d_{j,l}}) - \left( {{r^d_{j,k}} + {r^d_{i,l}}} \right) \ge 0.$ We collect all such Kalmanson inequality constraints in the set $\mathcal{K}$. Since all elements of $\mathcal{K}$ must be non-negative, we denote this condition as $\mathcal{K} \succeq 0$, representing all feasible Kalmanson inequality conditions for resistance distances in CPPR networks. {Both, $\Delta \succeq 0$ and $\mathcal{K} \succeq 0$ are utilized in optimization formulation $\mathbf{\Pi}_1$, $\mathbf{\Pi}_2$ and $\mathbf{\Pi}_3$.} In the next section, we present the network initialization method, which constitutes the first stage of our multi-stage topology reconstruction approach. 
\section{Network Initialization}\label{sec:initalnetwork}
\subsection{Construction of $MPRSN$}
\begin{figure}
	\centering
	\includegraphics[scale=0.15]{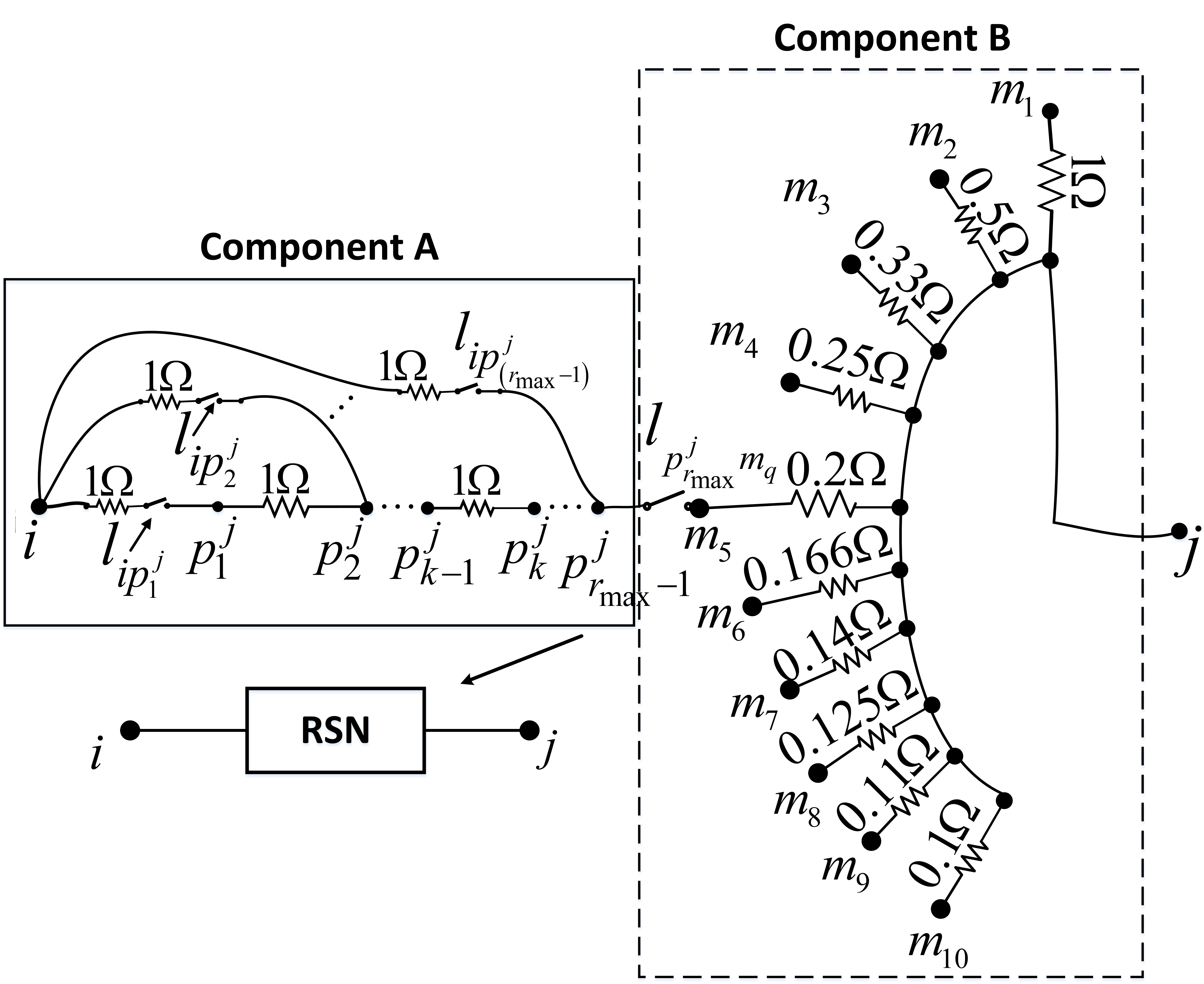}
	\caption{General construction of a resistor-switch network.}
	\label{fig:gen_RSN}
\end{figure}

Since the structure of the unknown network $\Gamma$ is not known apriori, we initiate the topology reconstruction process by constructing a \emph{maximal planar resistor-switch network} (MPRSN) over the $n_b$ boundary nodes. The MPRSN is an electrical network formed by embedding resistors and switches on each edge of a maximal circular planar graph.
To construct the MPRSN, we first generate a \emph{maximal circular planar graph} using the $n_b$ boundary nodes. A graph is considered maximal in this context if adding any additional edge would render it non-planar. Let $\mathcal{G}^{max}_{n_b} = (\mathcal{V_B}, \mathcal{E}^{max})$ denote this maximal planar graph defined on the $n_b$ boundary nodes. Formally, we define it as follows:
\begin{defn}\label{max_planar}(Maximal circular planar graph)
$\mathcal{G}^{max}_{n_b}$ is said to be a \emph{maximal circular planar graph} on $n_b$ boundary nodes if,
\begin{enumerate*}
\item[\textbf{1)}] it has $n_b$ boundary nodes arranged in a circular clockwise direction on circle $C$,
\item[\textbf{2)}] On $n_b$ boundary nodes we construct a graph with $3n_b-6$ edges, which is a maximal planar graph \cite{nishizeki2004planar}.
\end{enumerate*}
\end{defn} 
\noindent The construction of the maximal planar resistor-switch network (MPRSN) is carried out in three systematic steps, as outlined below:
\begin{enumerate}
    \item In the first step, we construct a maximal circular planar graph, denoted as $\mathcal{G}^{max}_{n_b}$, using the $n_b$ boundary nodes.
    \item In the second step, we design a network composed of interconnected resistors and switches, referred to as a resistor-switch network (RSN). The number of resistors and switches in each RSN is determined based on the parameter $r_{max} = \gamma_{min}^{-1}$. By appropriately configuring the ``on'' and ``off'' states of the switches, the RSN can generate a range of resistance values along its edges.
    \item Finally, each edge in the maximal circular planar graph $\mathcal{G}^{max}_{n_b}$ is replaced with a corresponding RSN. This substitution results in the complete construction of the MPRSN, which serves as the foundation for subsequent stages of the topology reconstruction process. 
\end{enumerate}
We now provide a concise explanation of each step involved in constructing the maximal planar resistor-switch network (MPRSN). In the first step, we construct a circular graph with $3n_b - 6$ edges on $n_b$ boundary nodes. This ensures maximal planarity, as defined in Definition \ref{max_planar}. In the second step, we design a resistor-switch network (RSN) based on the parameter $r_{max} = \gamma_{min}^{-1} \in \mathbb{R}^+$, as illustrated in Fig. \ref{fig:gen_RSN}. We denote this general RSN between boundary nodes $i$ and $j$ as $\mathcal{C}_{ij}$, and let $\mathcal{C} = \{\mathcal{C}_{ij} : ij \in \mathcal{E}^{max}\}$ represent the collection of all such RSN corresponding to the edges of the maximal planar graph $\mathcal{G}^{max}_{n_b}$. Each $\mathcal{C}_{ij}$ consists of two components: Component A and Component B, as shown in Fig. \ref{fig:gen_RSN}. These components are designed to approximately generate all admissible values of edge resistance $r(ij) \leq r_{max}$ by appropriately configuring the switches. A brief explanation of these components follows.
\subsubsection{Component A}
Component A of $\mathcal{C}_{ij}$ consists of a boundary node $i$, intermediate nodes $p_s^j$, and their corresponding switch variables $l_{ip_s^j}$, where $s \in \mathbf{Z}^+_{\leq (r_{max}-1)}$. Each switch variable can take a value of either 0 or 1, resulting in $2^{(r_{max}-1)}$ possible switch combinations. These combinations induce $2^{(r_{max}-1)}$ distinct resistance values for $r(ip^j_{r_{max}-1})$. The minimum resistance value, $r_{min}(ip^j_{r_{max}-1})$, is achieved when all switches are turned on, while the maximum resistance value is $r_{max}-1$. However, component A can only generate $2^{(r_{max}-1)}$ discrete resistance values within the range $[0, r_{max}-1]$, limiting its resolution and ability to produce other intermediate values. To address this limitation, we introduce an additional component, referred to as component B, as illustrated in Figure \ref{fig:gen_RSN}.
\subsubsection{Component B}
Component B of $\mathcal{C}_{ij}$ is designed to generate fractional edge resistance values $r(m_qj)$, where $q \in \mathbf{Z}^+_{\leq 10}$. It comprises 10 resistances arranged such that each edge resistance $r(m_qj)$ corresponds to the parallel combination of $q$ resistors, each with a resistance of $1\Omega$, as shown in Figure \ref{fig:gen_RSN}. Component B is connected to component A via a switch $l_{p^j_{(r_{max}-1)}m_q}$. Together, components A and B enable the resistor-switch network $\mathcal{C}_{ij}$ to approximately generate any resistance value within the range $[r_{min}(ip^j_{r_{max}-1}) + 0.1, r_{max}]$. Alternative designs could further enhance the range and precision of generated resistance values within $[0, r_{max}]$.
To construct the maximal planar resistor-switch network (MPRSN), each edge $ij \in \mathcal{E}^{max}$ in the maximal circular planar graph $\mathcal{G}^{max}_{n_b}$ is replaced by a corresponding resistor-switch network $\mathcal{C}_{ij}$. This completes the construction of the MPRSN. For a more detailed understanding, an example of constructing an MPRSN is provided in Example \ref{exam_MPRSN}.
\begin{rem}
The number of edges in an unknown CPPR network $\Gamma$ is bounded above by $3n_b - 6$, as per Definition \ref{max_planar}. To identify the edges in the unknown topology, a switching-based structure is used. The switch $l_{p^j_{(r_{max}-1)}m_q}$ determines the presence of a specific edge $ij$ based on resistance distance measurements and the Kirchhoff index.
\end{rem}
\begin{exmp}\label{exam_MPRSN}
Let $n_b=4$ and $r_{max}=\gamma_{min}^{-1}=4\Omega$ are known apriori. The first step is the construction of $\mathcal{G}^{max}_4$, on $4$ boundary nodes, as shown in Fig.\ref{fig:maxplangraph}.  
	\begin{figure}[h]
		\centering
		\includegraphics[scale=0.2]{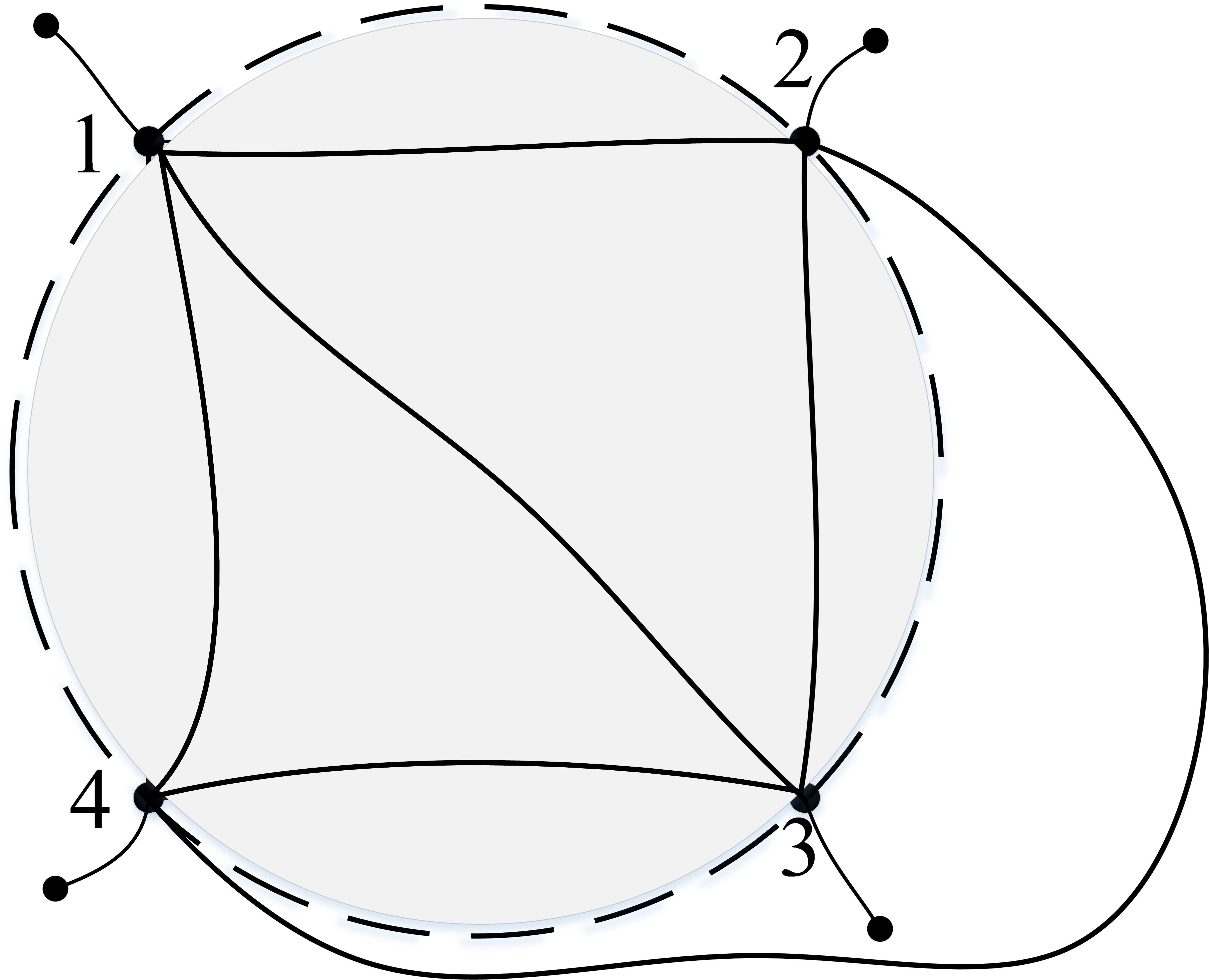}
		\caption{Maximal circular planar graph \texorpdfstring{$\mathcal{G}^{max}_4$}{Gmax4}.}
		\label{fig:maxplangraph}
	\end{figure}
	\noindent The second step is constructing the resistor switch network $\mathcal{C}_{ij}$ across the boundary nodes $i,j \in \mathcal{V_B}$, based on $r_{max}$, as shown in Fig.\ref{exmprsn}. 
	\begin{figure}[h]
		\centering
\includegraphics[width=0.25\textwidth]{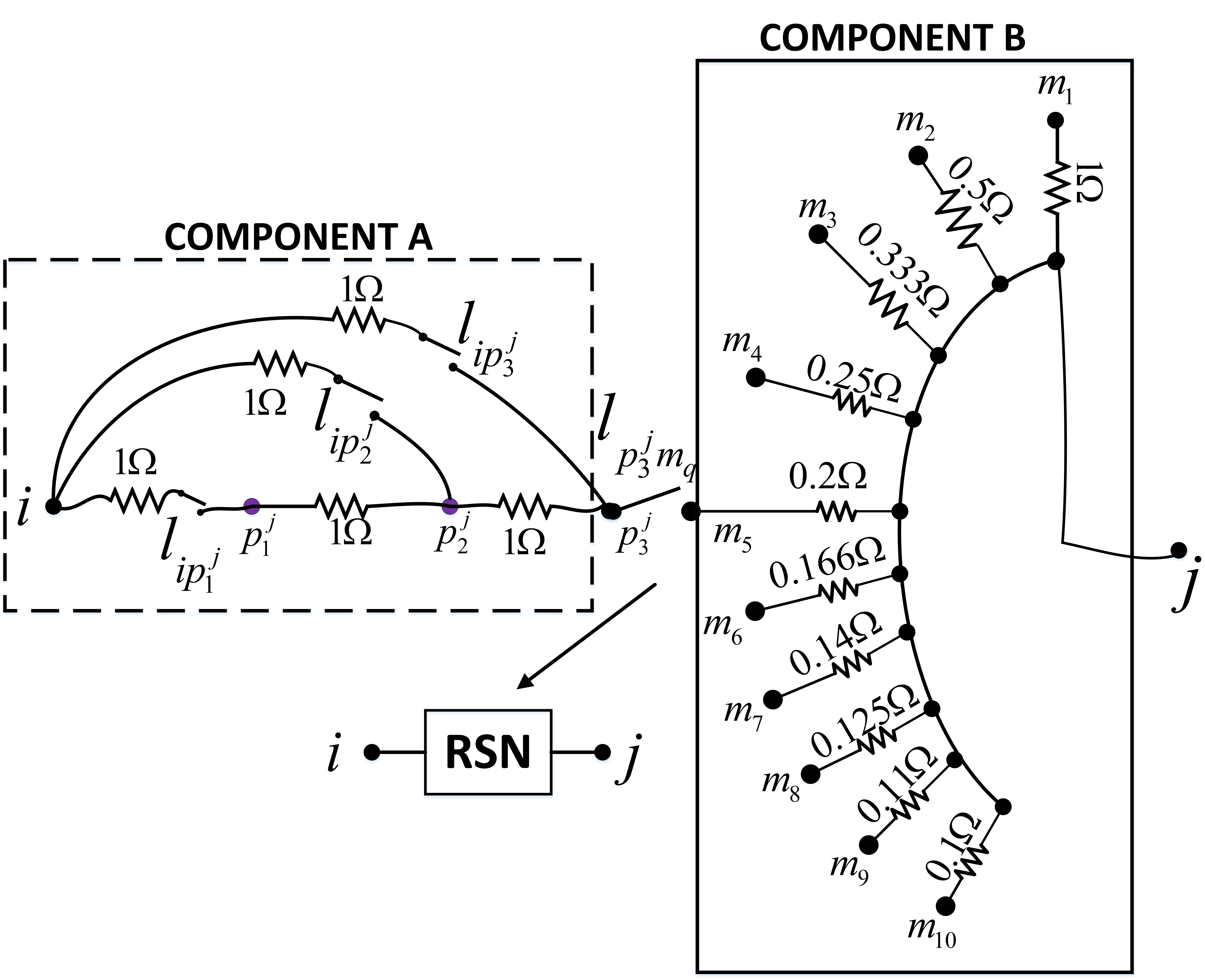}
		\caption{Resistor switch network.}
		\label{exmprsn}
	\end{figure}
	Component A of $\mathcal{C}_{ij}$ has three switches $l_{ip^j_1}, l_{ip^j_2}, l_{ip^j_3}$, therefore, there are $2^3$ switch combinations that induce $2^3$ resistance values. These resistance values are $\{\infty, 1, 2, 0.666, 3, 0.75, 1.66, 0.625\}$. The minimum value $0.625\Omega$ is obtained when all switches in component A are $on$, and the maximum value is $3\Omega$ other than $\infty \Omega$. Component B is added to component $A$ through a switch $l_{p_3^jm_q}$, where $1\le q \le 10$ and $q \in \mathbf{Z}^+_{\le 10}$. The designed $\mathcal{C}_{ij}$, approximately generates resistance values $r\left(ij\right)$ in the range $\left[0.725\,\,4\right]$ ($\infty$ not included). \\
In the last step, we replace each edge in $\mathcal{G}^{max}_4$ by a resistor switch network. The resultant $MPRSN$ is as shown in Fig.\ref{fig:mprsn}.
	\begin{figure}
		\centering
\includegraphics[width=0.15\textwidth]{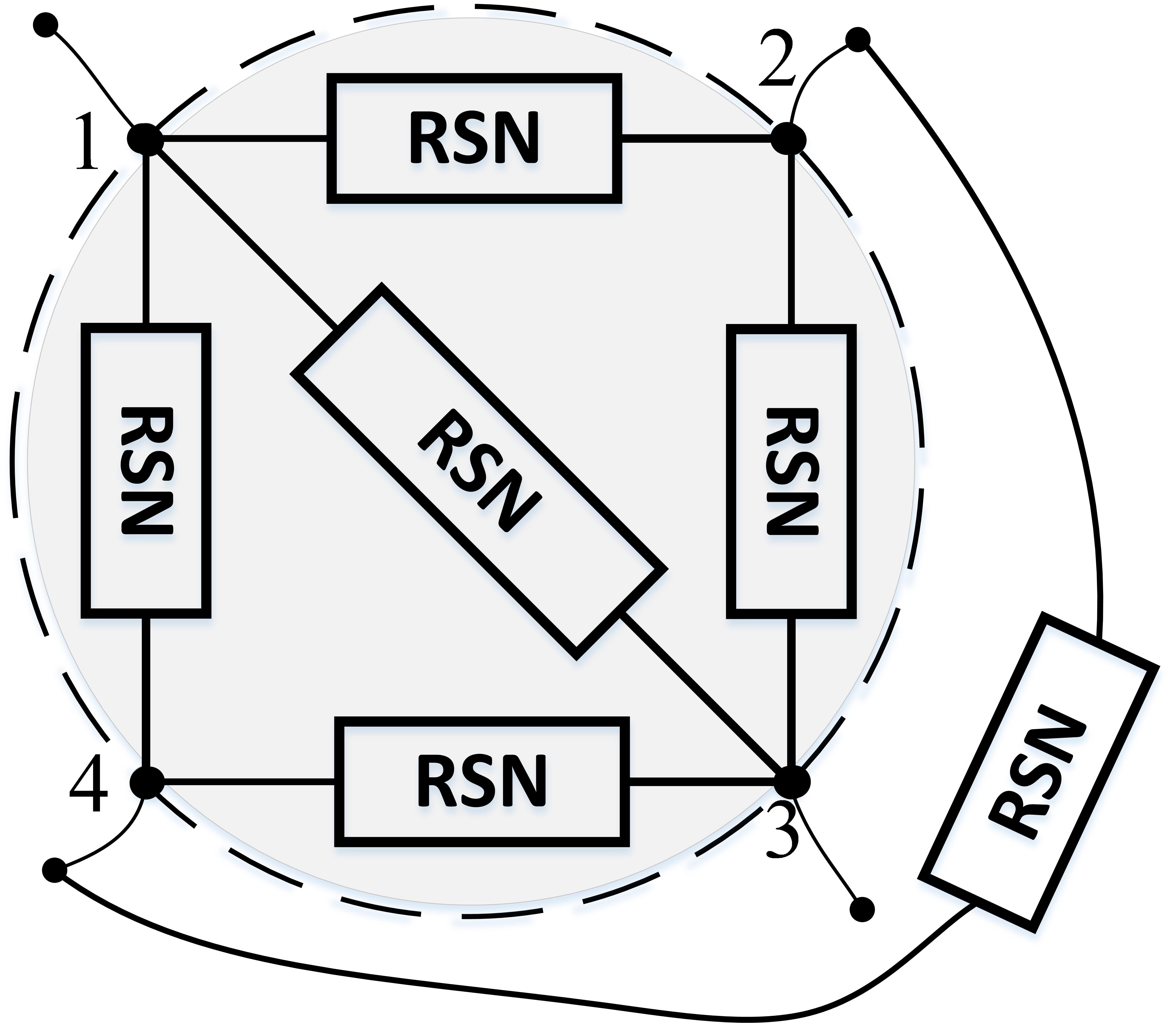}
		\caption{MPRSN on 4 boundary nodes}
		\label{fig:mprsn}
	\end{figure}  
\end{exmp}
Let the constructed $MPRSN$ be called as $\Gamma_{M}=\left(\mathcal{G}_{M},\gamma_M \right)$. Where $\mathcal{G}_{M}=\left( \mathcal{V}_M,\mathcal{E}_M\right)$, $\mathcal{V}_M$ is the set of all nodes and $\mathcal{E}_M$ is a set of all edge in $\Gamma_M$. Let $\mathbf{S}_M \subset \mathcal{E}_M$ is a set of all node pairs connected through a resistor and switch, for example, in Fig.\ref{exmprsn}, a node pair $i$ and $p_1^j$ are connected by a $1\Omega$ resistor and a switch, therefore $ip_1^j \in \mathbf{S}_M$. The conductance function $\gamma_M:\mathcal{C}\rightarrow\mathbf{R}^+$. The Laplacian matrix of $\Gamma_M$ is $\mathcal{L}\left(\Gamma_M\right)$, defined as follows. 
\begin{equation}
\left[\mathcal{L}\left(\Gamma_M\right)\right]_{kl}=\left[\mathcal{L}_{kl}\right]=\left\{
	\begin{array}{ll}
		-\gamma\left(kl\right)=-1, & \mbox{if $kl\in \mathcal{E}_M \setminus \mathbf{S}_M$},\\
		-\gamma\left(kl\right)=-l_{kl}, & \mbox{if $kl\in \mathbf{S}_M$},\\
		\sum\limits_{l\in \mathcal{N}\left( k \right)}{{\gamma\left(kl\right)}}, & \mbox{if $k=l$},\\
		0, & \mbox{otherwise}.
	\end{array}
	\right.
\end{equation}   
\noindent The Laplacian matrix $\mathcal{L}(\Gamma_M)$ of the maximal planar resistor-switch network (MPRSN) is significantly larger than $\mathcal{L}(\Gamma)$ of the unknown network. In $\Gamma_M$, the goal is to find an optimal combination of switch positions that minimizes the difference between measured resistance distances $r^d_{i,j}$ and those in $\Gamma_M$, $r^d_{i,j}(\Gamma_M)$, for all boundary node pairs $i, j \in \mathcal{A}$. Here, $r^d_{i,j}(\Gamma_M)$ depends on the switch positions. In the next section, we formulate an optimization problem, $\mathbf{\Pi}_1$, to determine these optimal switch positions, ensuring the reconstructed network aligns closely with the available measurements.
\subsection{Determining Switch Positions in $\Gamma_M$}
\label{sec:const_auxgamma}
The switch position variables $l_{ip_s^j}$ within each resistor-switch network $\mathcal{C}_{ij}$ are initially unknown. To determine these switch positions, we define a binary vector $\bm{\rho} \in \{0,1\}^t$, where $t = (3n_b - 6)(\lfloor r_{max} \rfloor - 1) + 10$, representing the states of all $t$ switches in the network. To compute $\bm{\rho}$, we formulate two optimization problems: $\mathcal{I}$ and $\mathbf{\Pi_1}$. The first problem, $\mathcal{I}$, is designed to estimate the unknown resistance distances $r^d_{i,j}$ for all node pairs $i,j \in \mathcal{U_B}$ (the unavailable boundary nodes). These estimates are denoted as $\hat{r}^d_{i,j}\,\forall i,j \in \mathcal{U_B}$. The second problem, $\mathbf{\Pi_1}$, utilizes both the known resistance distances $r^d_{i,j}\,\forall i,j \in \mathcal{A}$ (the available boundary nodes) and the estimated resistance distances $\hat{r}^d_{i,j}\,\forall i,j \in \mathcal{U_B}$ to determine the optimal switch positions. Below, we first present the formulation of problem $\mathcal{I}$, followed by a detailed explanation of $\mathbf{\Pi_1}$.
\subsubsection{Optimization problem $\mathcal{I}$}
Consider resistance distance equation (\ref{eqn:resis_equa}), which can be expressed as, $r^d_{s,t} = x_{ss} + x_{tt} - 2x_{st}, \forall s,t \in \mathcal{A},$ where $x_{st}=X(s,t)$. This formulation yields  $\frac{|\mathcal{A}|\left(|\mathcal{A}|+1\right)}{2}$ linear equations. Additionally, from the definition of the matrix $\textbf{X}$, we know that $\textbf{X}\mathbf{1}=\mathbf{1}$, which provides $m=n_b+n_i$ linear equations. Furthermore, since  Kirchhoffs index $K_{\Gamma}$ is known, equation (\ref{eqn:kirchhoffindex}) is incorporated as an additional linear constraint. By combining these equations, we obtain a system of $\frac{|\mathcal{A}|\left(|\mathcal{A}|+1\right)}{2}+m+1$ linear equations, represented compactly as:
\begin{equation}\label{eqn:laplstru}
	\textbf{A}\mathbf{x}=\begin{bmatrix} \mathbf{r} \\ \mathbf{1} \end{bmatrix}.
\end{equation}
Where $\mathbf{x} \in \mathbf{R}^{\frac{m(m+1)}{2} \times 1}$ is the vector of unknowns derived from the symmetric matrix $\textbf{X} \in \mathcal{S}_m$, and $\mathbf{r} \in \mathbf{R}^{\frac{|\mathcal{A}|\left(|\mathcal{A}|-1\right)+2}{2}}$ is the vector of known resistance distances, augmented with the value $K_{\Gamma}$. To estimates the unknown resistance distances $r^d_{i,j} \forall i,j \in \mathcal{U_B}$, we solve the following optimization problem: 
\begin{equation*}\label{initaloptimiformulation}
		\min_{\mathbf{x}} \,  \norm{\textbf{A}\mathbf{x}-\begin{bmatrix} \mathbf{r} \\ \mathbf{1} \end{bmatrix}}_2^2\\
		\textrm{s.t}\,\,  \mathcal{T}\left(x\right)\succ 0,
		\Delta \succeq 0,
		\mathcal{K} \succeq 0. \hspace{1.6cm} \left(\mathcal{I}\right)
\end{equation*}
where $\mathcal{T}: \mathbb{R}^{\frac{m(m+1)}{2} \times 1} \to \mathcal{S}_m$ is a transformation mapping $\mathbf{x}$ to the symmetric matrix $\textbf{X}$. The constraints $\Delta \succeq 0$ and $\mathcal{K} \succeq 0$ ensure that the triangle inequality and Kalmanson’s inequalities are satisfied, respectively. {The optimization problem $\mathcal{I}$ is convex but non-strict and may admit multiple solutions due to the underdetermined system. However, the goal is not to find a unique solution but to estimate valid resistance distance measurements consistent with the physical and mathematical properties of a CPPR network.}

The solution to problem $\mathcal{I}$ is denoted as $\hat{\mathbf{x}}$. From $\hat{\mathbf{x}}$, we construct the estimated matrix $\hat{\textbf{X}}$ using $\mathcal{T}(\hat{\mathbf{x}}) = \hat{\textbf{X}}$. Subsequently, the estimated resistance distance matrix $\hat{\textbf{R}}_{\Gamma}$ is computed using the relation (\ref{eqn:resis_equa}). The entries of $\hat{\textbf{R}}_{\Gamma}$ provide the estimated resistance distances: $\hat{r}^d_{i,j} = \hat{R}_{\Gamma}(i, j), \quad \forall i, j \in \mathcal{U_B}.$
\subsubsection{Optimization problem $\mathbf{\Pi_1}$}
The optimization problem $\mathbf{\Pi_1}$ is formulated to minimize the resistance distance error, defined as: $\tilde{r}_{i,j}^d = 
\begin{cases} 
r_{i,j}^d - r_{i,j}^d(\Gamma_M), & \text{if } i,j \in \mathcal{A}, \\
\hat{r}_{i,j}^d - r_{i,j}^d(\Gamma_M), & \text{if } i,j \in \mathcal{U_B},
\end{cases}
\quad \forall i,j \in \mathcal{V_B}.$
Here, $r_{i,j}^d \in r^d$ represents the measured resistance distances for node pairs in $\mathcal{A}$, while $\hat{r}_{i,j}^d$ denotes the estimated resistance distances for node pairs in $\mathcal{U_B}$. The term $r_{i,j}^d(\Gamma_M)$ corresponds to the resistance distance across boundary nodes $i,j$ in the maximal planar resistor-switch network ($\Gamma_M$), which depends on the switch positions. To proceed, we define $\textbf{r}^{d*} \in \mathbb{R}^{\frac{n_b(n_b-1)}{2} \times 1}$ as the vector of measured and estimated resistance distances, and $\textbf{r}^{d}(\Gamma_M) \in \mathbb{R}^{\frac{n_b(n_b-1)}{2} \times 1}$ as the vector of resistance distances in $\Gamma_M$, which is a function of the switch positions $\bm{\rho}$. The resistance distance error vector is then expressed as: $\tilde{\textbf{r}}^d \triangleq \textbf{r}^{d*} - \textbf{r}^{d}(\Gamma_M).$ The objective of $\mathbf{\Pi_1}$ is to minimize this error vector with respect to the switch positions $\bm{\rho}$. The formulation of $\mathbf{\Pi_1}$ is presented below.
\begin{equation*}\label{optimiformulation}
	\begin{aligned}
	     \min_{\bm{\rho},\textbf{W}} \, & ({\tilde{\textbf{r}}^d})^T \textbf{W}  {\tilde{\textbf{r}}^d}\hspace{3.2cm} \left(\mathbf{\Pi_1}\right) \\
		\textrm{s.t} \,\, & \bm{\rho} \odot \left( \mathbf{1}-\bm{\rho}\right){\succeq \mathbf{0}}, \Delta \succeq 0,
		\mathcal{K} \succeq 0,
  0.5 \le W_{ij}\le 0.9\, \forall i,j \in \mathcal{U_B}
	\end{aligned}
\end{equation*}
In the optimization problem $\mathbf{\Pi_1}$, the weighting matrix $\textbf{W} \in \mathbb{R}^{\frac{n_b(n_b-1)}{2} \times \frac{n_b(n_b-1)}{2}}$ is a diagonal matrix with positive entries. The diagonal elements $W_{ij}$ weight the resistance distance errors $\tilde{r}_{i,j}^d$. For node pairs $i,j \in \mathcal{A}$ (available boundary nodes), $W_{ij}$ is fixed to $1$, while for $i,j \in \mathcal{U_B}$ (unavailable boundary nodes), $W_{ij}$ is constrained to the range $[0.5, 0.9]$, as defined in $\mathbf{\Pi_1}$. {The binary constraints on $\bm{\rho} \in \{0, 1\}$ introduce non-convexity, complicating the optimization. To address this, $\bm{\rho}$ is relaxed to the interval $[0, 1]$.} Additionally, the constraints involving terms $\left( r_{i,k}^d + r_{j,l}^d \right) - \left( r_{i,j}^d + r_{k,l}^d \right)$ and $\left( r_{i,k}^d + r_{j,l}^d \right) - \left( r_{j,k}^d + r_{i,l}^d \right)$ in $\mathcal{K}$ represent differences of convex functions. Consequently, $\mathbf{\Pi_1}$ is formulated as a difference of convex programming (DCCP) problem \cite{shen2016disciplined}. To solve $\mathbf{\Pi_1}$, we employ the \emph{disciplined convex concave programming} package \cite{shen2016disciplined, lipp2016variations}, which is specifically designed for such problems. While various methods for computing initial guesses are discussed in \cite{shen2016disciplined}, we propose a novel approach to construct an initial guess for $\mathbf{\Pi_1}$, detailed in Appendix \ref{append:choice_initial}.\\
A term $({\tilde{\textbf{r}}^d})^T \textbf{W}  {\tilde{\textbf{r}}^d}$ in objective function is convex if and only if each $r^d_{i,j}\left( \Gamma_M \right)$ is convex with respect to the edge conductances. The convexity of resistance distance with respect to the edge conductance is discussed in \cite{ghosh2008minimizing}. Here, we mention the same 
as proposition \ref{convexrho}.
\begin{prop}\label{convexrho}
Let $\mathbf{c}$ be a vector of edge conductances of any $\Gamma$. The resistance distance $r_{s,t}^{d}\left(\mathbf{c}\right)$ is a convex function of $\mathbf{c}$. 
\end{prop}
\noindent The solution to $\mathbf{\Pi_1}$ yields a vector $\bm{\rho} \in [0,1]^p$. However, since the elements of $\bm{\rho}$ must be binary (either $0$ or $1$), we apply the \emph{Round-Down algorithm} \cite{boros2002pseudo} to convert the continuous values into a boolean vector. 
\subsubsection{Round Down Algorithm}
Constraining $\bm{\rho}$ to binary values (either $0$ or $1$) introduces non-convex constraints, which complicates the optimization process. To address this, $\bm{\rho}$ is relaxed to lie within the interval $[0, 1]$. The resulting solution vector $\bm{\rho} \in [0, 1]^t$ is then converted into a boolean vector $\mathbf{x} \in \{0, 1\}^t$ using the Round-Down algorithm \cite{boros2002pseudo}. This algorithm is grounded in Proposition \ref{prop:bool}, stated below:
\begin{prop}\cite{boros2002pseudo}\label{prop:bool}
Consider a boolean function $f: \{0, 1\}^n \to \mathbb{R}$ and let $\bm{\rho} \in \mathbb{R}^n$. There exist boolean vectors $\mathbf{x}, \mathbf{y} \in \{0, 1\}^n$ such that $f(\mathbf{x}) \leq f(\bm{\rho}) \leq f(\mathbf{y})$.
\end{prop}
\noindent Based on this proposition, there exists a boolean vector $\mathbf{x}$ that minimizes the objective function $f_o(\bm{\rho})$ in $\mathbf{\Pi_1}$. To compute $\mathbf{x}$ efficiently, the derivative of $f_o(\bm{\rho})$, denoted as $\delta_i(\bm{\rho})$, is defined for each element $\rho_i$ as follows:
\begin{equation}\label{boolean_derivative}
\delta_i(\bm{\rho}) \triangleq f_o(\ldots, \rho_{i-1}, 1, \rho_{i+1}, \ldots) - f_o(\ldots, \rho_{i-1}, 0, \rho_{i+1}, \ldots).
\end{equation}
\noindent The Round-Down algorithm evaluates each element $\rho_i \in (0, 1)$ in the solution vector $\bm{\rho}$. For such elements, it computes the derivative $\delta_i(\bm{\rho})$. Depending on the sign of $\delta_i(\bm{\rho})$, the $i^{th}$ element is set to either $0$ or $1$. After flipping the value, the algorithm verifies whether the triangle inequality and Kalmanson constraints are satisfied. A detailed description of the Round-Down algorithm is provided in Algorithm \ref{euclid}.\\
\begin{algorithm}
	\caption{Round-Down Algorithm}\label{euclid}
	\begin{algorithmic}[1]
		\Require $i \gets 1$, $\mathbf{q}^0 \gets \bm{\rho}$ \& $k \leftarrow 0$
		\Repeat
		\State $k \leftarrow k+1$
		\If{$0 < q_i^{(k-1)} < 1$ \& $\delta_i\left(\mathbf{q}^{(k-1)}\right) > 0$}
		\State $q_i^{(k)} \leftarrow 0$ 
        \If{$\Delta\left(\mathbf{q}^{(k-1)}\right) \ge 0$ \& $\mathcal{K}\left(\mathbf{q}^{(k-1)}\right) \ge 0$}
        \State $q_i^{(k)} \leftarrow 0$ 
        \Else 
        \State $q_i^{(k)} \leftarrow 1$ 
         \EndIf
		\ElsIf{$0 < q_i^{(k)} < 1$ \& $\delta_i\left(\mathbf{q}^{(k-1)}\right) < 0$}
        \State $q_i^{(k)} \leftarrow 1$ 
		\If{$\Delta\left(\mathbf{q}^{(k-1)}\right) \ge 0$ \& $\mathcal{K}\left(\mathbf{q}^{(k-1)}\right) \ge 0$}
        \State $q_i^{(k)} \leftarrow 1$ 
        \Else 
        \State $q_i^{(k)} \leftarrow 0$ 
         \EndIf
		\Else 
		\State $q_i^{(k)} \leftarrow q_i^{(k-1)}$
		\EndIf
		\State $i \leftarrow i+1$
		\Until{$i \le n$}
        \State x=q
	\end{algorithmic}
\end{algorithm}
The boolean vector $\mathbf{x}$ specifies the optimal switch positions, which are then used to construct an initial resistive network. We refer to this network as the auxiliary network, denoted by $\Gamma_{aux} = (\mathcal{G}_{aux}, \gamma_{aux})$, where $\mathcal{G}_{aux} = (\mathcal{V_B}, \mathcal{E}_{aux})$ represents the graph structure of the network. The auxiliary network $\Gamma_{aux}$ provides an initial topology $\mathcal{G}_{aux}$, which serves as the foundation for the subsequent stages of the reconstruction process. Additionally, the edge conductances $\gamma_{aux}: \mathcal{E}_{aux} \to \mathbb{R}^+$ are utilized as an initial guess in the optimization problem $\mathbf{\Pi}_2$ during later stages. Since $\Gamma_{aux}$ does not account for interior nodes, a systematic approach is required to incorporate them into the network. The detailed methodology for placing interior nodes within $\Gamma_{aux}$ is presented in Section \ref{place_int_nodes}.
\section{Placement of Interior Nodes}\label{place_int_nodes} 
We consider a modified network $\Bar{\Gamma}_{aux} = (\Bar{\mathcal{G}}_{aux}, \Bar{\gamma}_{aux})$, derived from the initial network $\Gamma_{aux}$ by replacing each edge conductance $\gamma_{aux}(\sigma)$ with an unknown variable $l_{\sigma}$, for all $\sigma \in \mathcal{E}_{aux}$. Here, $\Bar{\mathcal{G}}_{aux}$ retains the same structure as ${\mathcal{G}}_{aux}$, while the edge conductances are redefined as $\Bar{\gamma}_{aux}: \mathcal{E}_{aux} \to \mathbb{R}^+$. Let $\bar{\mathbf{c}} \in \mathbb{R}^{|\mathcal{E}_{aux}| \times 1}$ represent the vector of unknown edge conductances in $\Bar{\Gamma}_{aux}$, and let $\Bar{\textbf{r}}$ denote the corresponding edge resistance vector.

In this section, we aim to identify edges in $\Bar{\Gamma}_{aux}$ where interior nodes can be introduced. To achieve this, we formulate an optimization problem, denoted as $\mathbf{\Pi}_2$, and solve for the unknown edge conductances $\bar{\mathbf{c}}$. The problem $\mathbf{\Pi}_2$ is a reformulation of $\mathbf{\Pi}_1$, incorporating an additional term in the objective function that accounts for the error in the Kirchhoff index: $(K_{\Bar{\Gamma}_{aux}} - K_{\Gamma})^2$. Furthermore, the constraints on edge conductances are relaxed, as detailed below.
\begin{equation*}\label{optimiformulaux}
	\begin{aligned}
	     \min_{\bm{\rho},\textbf{W}} \, & ({\tilde{\textbf{r}}^d})^T \textbf{W}  {\tilde{\textbf{r}}^d} + \left( K_{\Bar{\Gamma}_{aux}}-K_{\Gamma} \right)^2		\hspace{3.2cm} \left(\mathbf{\Pi_2}\right)  \\
		\textrm{s.t} \,\, & \bar{\mathbf{c}}\,\succeq \mathbf{0}, \Delta \succeq 0,
		\mathcal{K} \succeq 0,
  0.5 \le W_{ij}\le 0.9\, \forall i,j \in \mathcal{U_B}.
	\end{aligned}
\end{equation*}
In the optimization problem $\mathbf{\Pi_2}$, $K_{\Bar{\Gamma}_{aux}}$ represents the Kirchhoff's index of the modified network $\Bar{\Gamma}_{aux}$, which is a function of the unknown edge conductances. The initial guess for the edge conductances, denoted as $\bar{\mathbf{c}}^{(0)}$, is derived from the edge conductances obtained in the initial network $\Gamma_{aux}$. The placement of interior nodes is determined based on the edge resistance vector $\Bar{\textbf{r}}$, which is the solution to $\mathbf{\Pi}_2$. To illustrate this, consider introducing an interior node $k$ on a resistive edge $ij \in \mathcal{E}_{aux}$. This operation splits the edge $ij$ into two new resistive edges, $ik$ and $kj$. According to the assumptions, each of these new edges can have a maximum resistance of $r_{max}$. However, in some cases, the total resistance of the original edge $ij$, given by $r(ij) = r(ik) + r(kj)$, may exceed $r_{max}$. To identify such edges, we examine the solution $\Bar{\textbf{r}}$ and extract all edge resistances greater than $r_{max}$. These resistances are then arranged in descending order and stored in a vector $\textbf{d}_r$. Let $n_{d_r}$ denote the number of elements in $\textbf{d}_r$. Based on the relationship between $n_{d_r}$ and the number of interior nodes $n_i$, the following rules are applied:
\begin{enumerate}
    \item If $n_{d_r} < n_i$, place $n_{d_r}$ interior nodes on the $n_{d_r}$ edges corresponding to the first $n_{d_r}$ entries in $\textbf{d}_r$. The remaining $n_i - n_{d_r}$ interior nodes are classified as dangling nodes (i.e., nodes with no incident edges).
    \item If $n_{d_r} > n_i$, place $n_i$ interior nodes on the $n_i$ edges corresponding to the first $n_i$ entries in $\textbf{d}_r$.
\end{enumerate}
\begin{figure*}
	\centering
	\subfloat[Unknown Network $\Gamma$]{\label{fig:unk_gamma}\includegraphics[width=.19\linewidth]{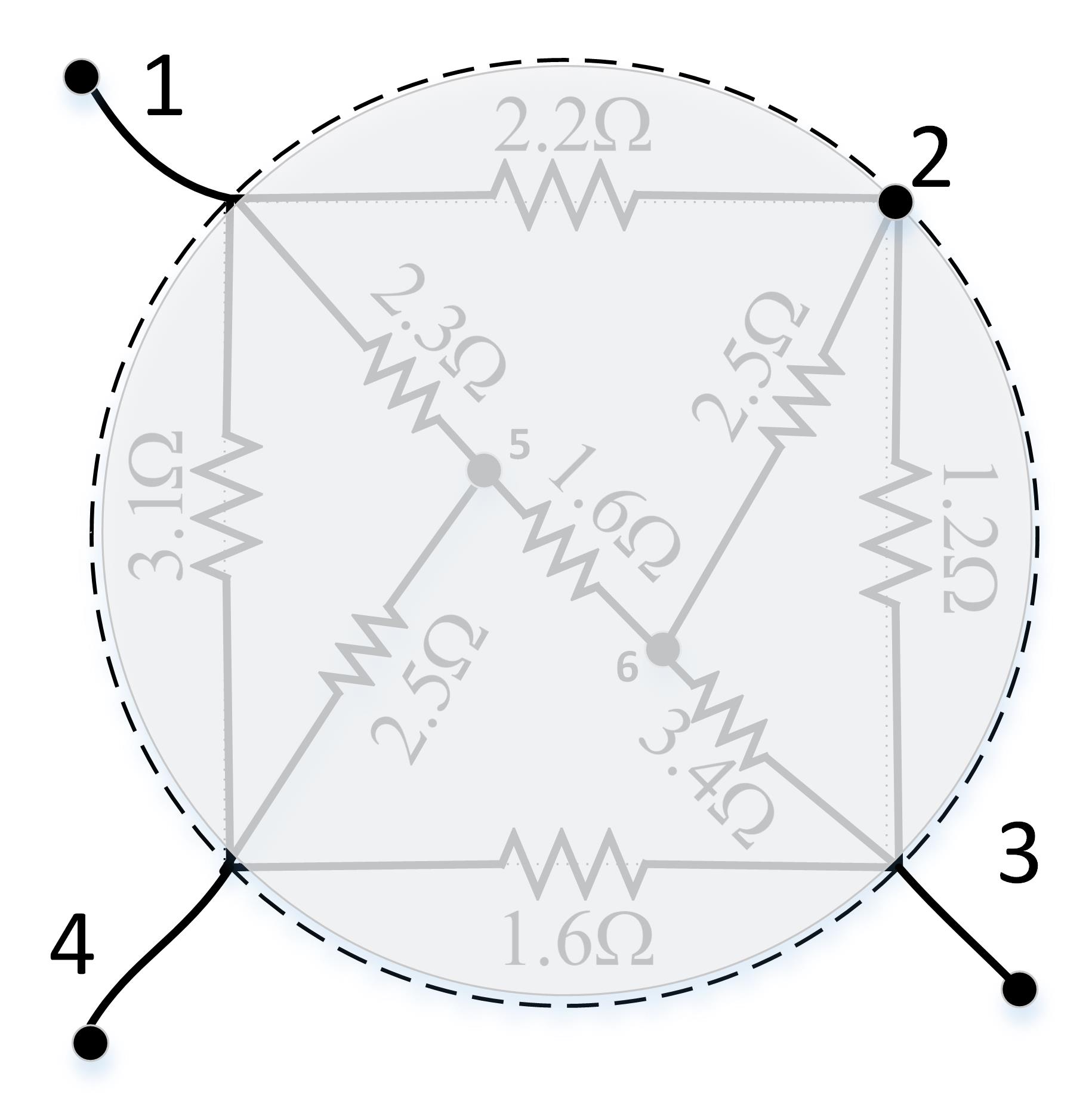}}\hfill
	\subfloat[Auxiliary network $\Gamma_{aux}$.] {\label{fig:aux_gamma}\includegraphics[width=.19\linewidth]{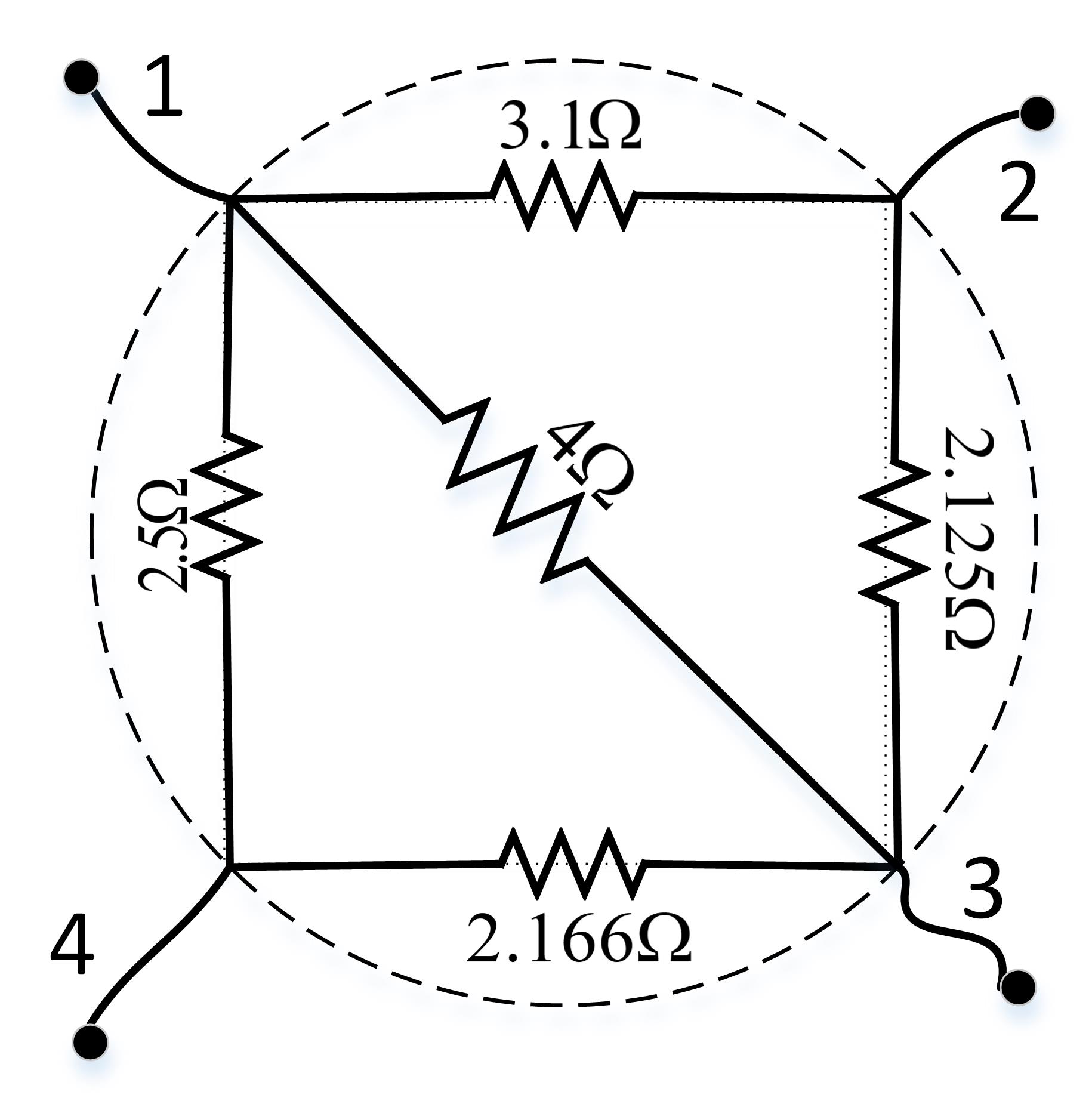}} \hfill
	\subfloat[$\bar{\Gamma}_{aux}$] {\label{fig:auxbar_gamma}\includegraphics[width=.19\linewidth]{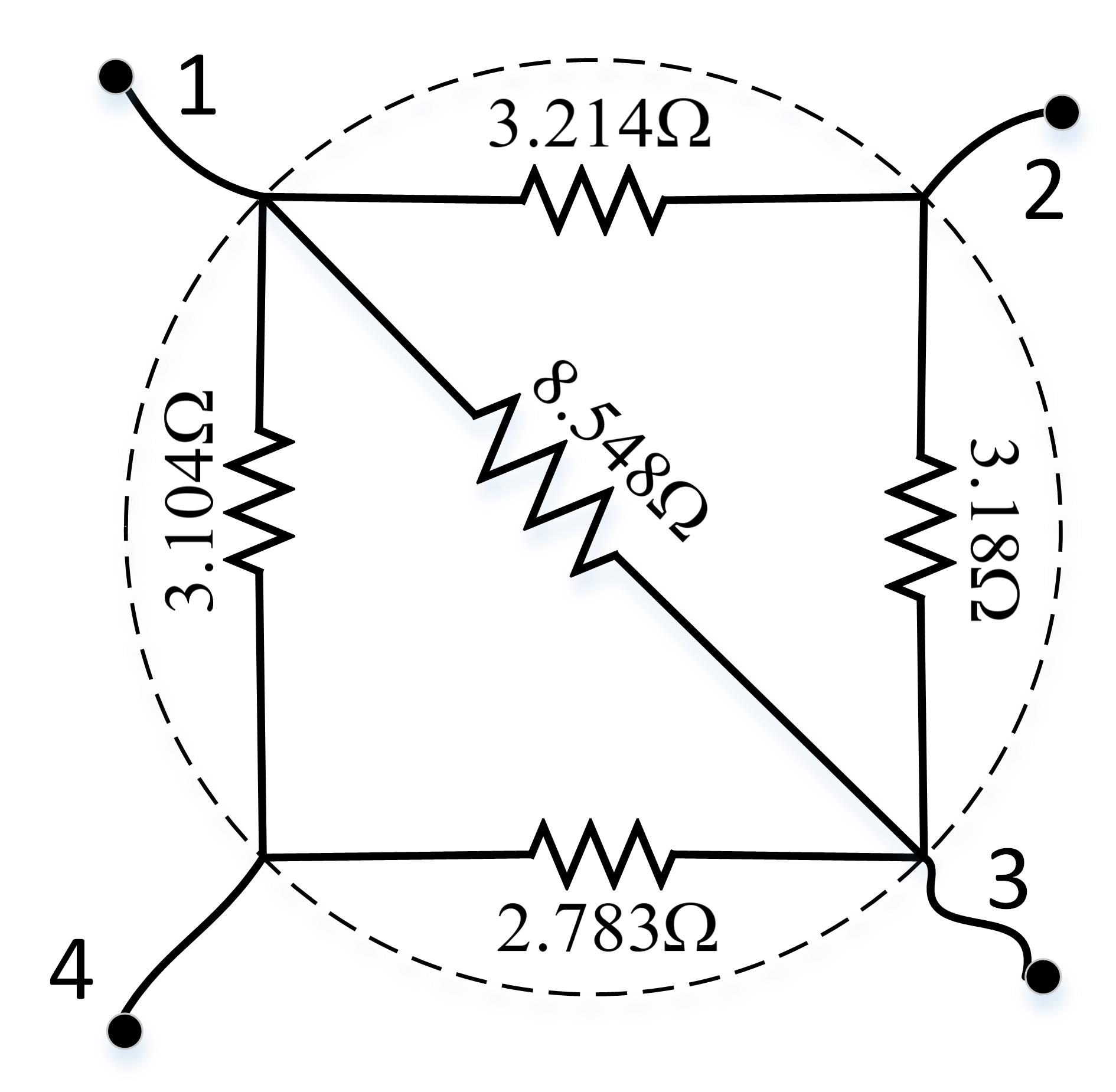}}\hfill
	\subfloat[$\hat{\Gamma}$]{\label{fig:auxhat_gamma}\includegraphics[width=.19\linewidth]{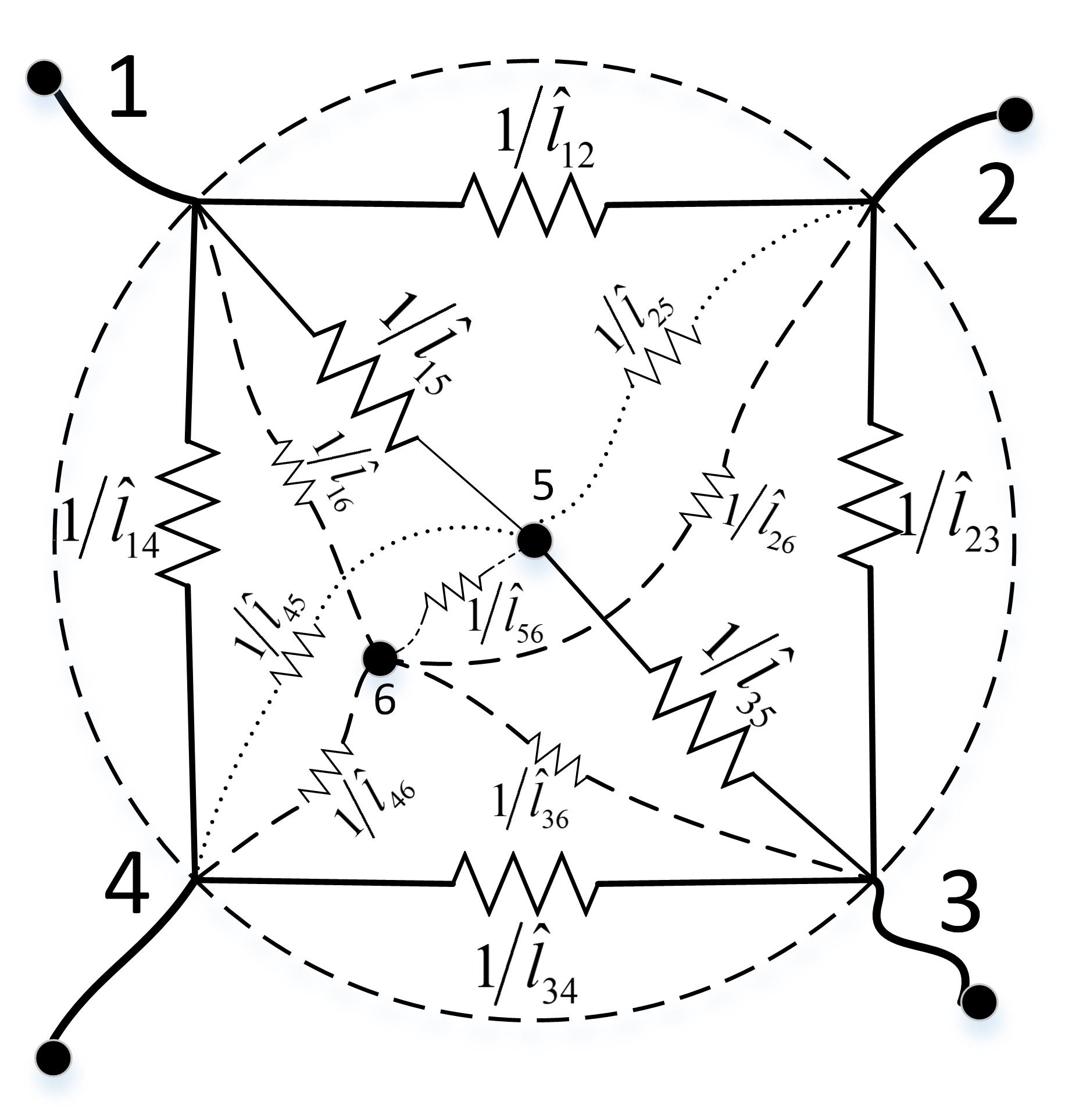}}
	\caption{Topology Reconstruction Example.}
\end{figure*}
Once the approximate positions of the $n_i$ interior nodes are determined, they are appropriately placed within the initial graph $\mathcal{G}_{aux}$. The next natural question arises: how are these interior nodes connected to the remaining nodes in the network? This question is addressed in detail in the following section.  
\section{Constructing Planar Networks and Rewiring}\label{cons_planar_net}
\subsection{Planarity checking and planar construction}
Once the approximate positions of the $n_i$ interior nodes are determined, they are appropriately placed within the initial graph $\mathcal{G}_{aux}$. The interior nodes are then connected to all other nodes in the network while preserving the existing edges in $\mathcal{E}_{aux}$. The resulting graph, denoted as $\hat{\mathcal{G}}_{aux} = (\mathcal{V_B}, \mathcal{V_I}, \hat{\mathcal{E}})$, includes the original edges $\mathcal{E}_{aux}$ and newly added edges $\mathcal{E}_p$, where $\mathcal{E}_p = \{ij : \forall i \in \mathcal{V_I} \text{ and } \forall j \in \mathcal{V_I} \cup \mathcal{V_B}\}$. However, connecting the interior nodes to all other nodes may render $\hat{\mathcal{G}}_{aux}$ non-planar. Since the goal is to reconstruct a planar resistive network, we extract a set of planar subgraphs from $\hat{\mathcal{G}}_{aux}$. To achieve this, we address two key questions: (1) How can we determine whether $\hat{\mathcal{G}}_{aux}$ is planar or non-planar? and (2) If the graph is non-planar, how can we extract planar subgraphs from it? These questions are addressed in the subsequent sections.
\subsubsection{Planarity Testing \& Construction} 
\begin{figure}[h]
	\centering
	\includegraphics[scale=0.25]{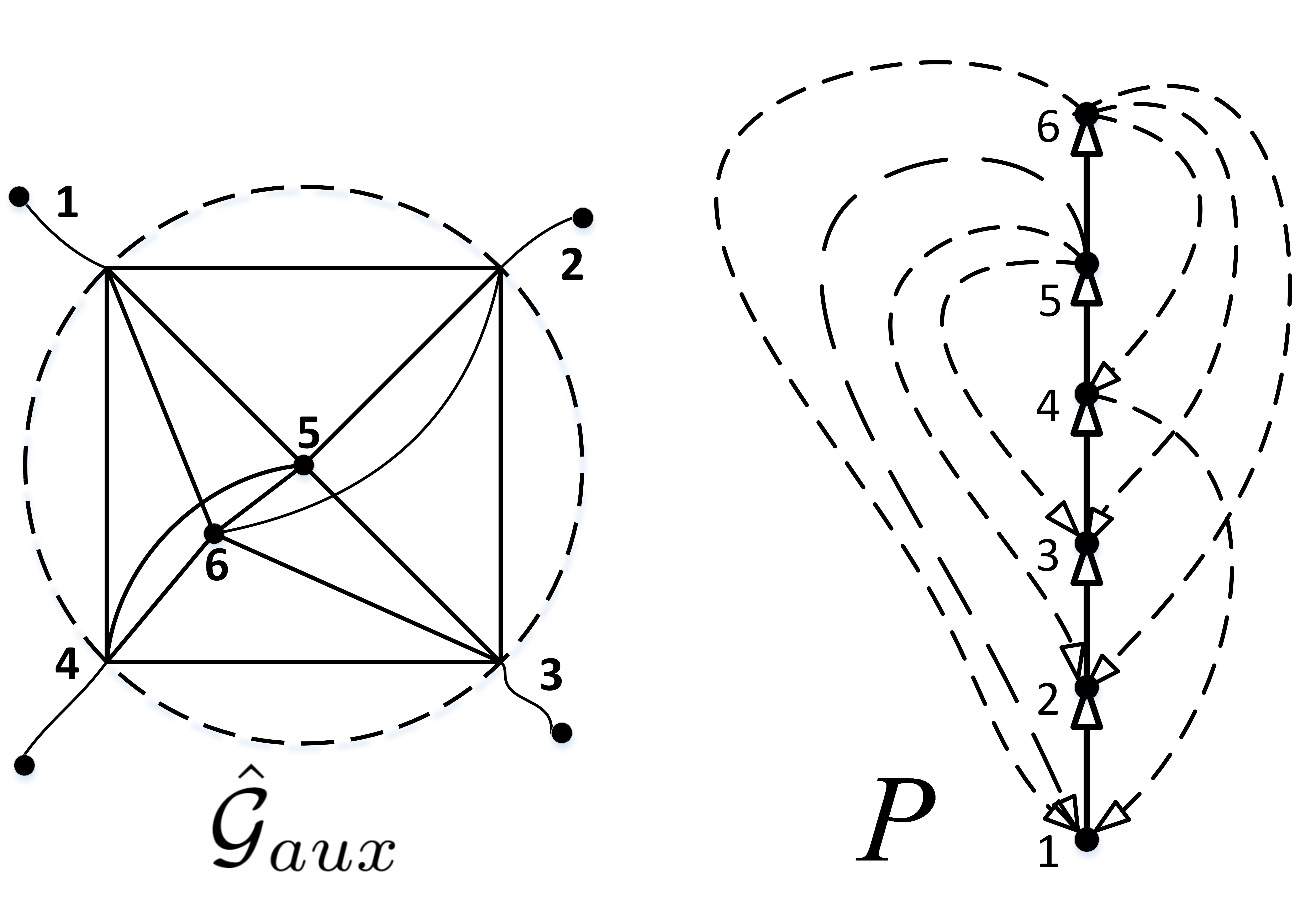}
	\caption{Palm tree representation $P$ of \texorpdfstring{$\hat{\mathcal{G}}_{aux}$}{Gaux}. Bold edges are the tree arcs, and dashed edges are the back edges.}
	\label{fig:palm}
\end{figure}

We define a transformation $\mathcal{T}$ that maps a graph $\hat{\mathcal{G}}_{aux}$ onto a plane with the following properties: 1. Each vertex in $\hat{\mathcal{G}}_{aux}$ is mapped to a distinct point on the plane. 2. Each edge in $\hat{\mathcal{G}}_{aux}$ is mapped to a simple curve on the plane, with its endpoints (vertices) mapped as specified in the first condition. The resulting diagram, $\mathcal{T}(\hat{\mathcal{G}}_{aux})$, is referred to as an embedding of $\hat{\mathcal{G}}_{aux}$. The graph $\hat{\mathcal{G}}_{aux}$ is considered planar if and only if no two distinct curves in $\mathcal{T}(\hat{\mathcal{G}}_{aux})$ intersect. In such cases, $\mathcal{T}(\hat{\mathcal{G}}_{aux})$ is called a planar embedding of $\hat{\mathcal{G}}_{aux}$ on the plane. Before presenting the planarity testing algorithm, we introduce key concepts necessary for understanding the process. To systematically explore an undirected graph $\hat{\mathcal{G}}_{aux}$, we employ the \emph{depth-first search (DFS)} algorithm. For a detailed explanation of DFS, refer to \cite{tarjan1972depth}. 
The DFS algorithm classifies edges into two categories: 1) \emph{Tree arcs}: Directed edges $ij$ (from $i$ to $j$) represented as $i \rightarrow j$, where $i < j$. 2) \emph{Back edges}: Directed edges $ij$ represented as $i \dashedrightarrow j$, where $i > j$.\\
\noindent If such partitions exist for $\hat{\mathcal{G}}_{aux}$, we construct a palm tree diagram $P$ to represent $\hat{\mathcal{G}}_{aux}$. An example of this palm tree representation is shown in Fig. \ref{fig:palm}, where bold edges represent tree arcs and dashed edges denote back edges. This representation corresponds to an example graph $\hat{\Gamma}$, as illustrated in Fig. \ref{fig:auxhat_gamma}.
\noindent To determine the planarity of $\hat{\mathcal{G}}_{aux}$, we first apply a depth-first search (DFS) to construct the palm tree representation $P$. Subsequently, we employ a modified version of the Auslander, Parter, and Goldstein algorithm \cite{auslander1961psv, goldstein1963efficient}. The algorithm operates as follows:
\begin{itemize}
\item \textbf{Cycle Identification and Deletion:} The algorithm identifies a cycle $c$ within the palm tree $P$ and removes it, resulting in a set of disconnected segments. An example of this process is shown in Fig. \ref{fig:palm_compo}(b) in \ref{append:mod_planar}.
\item \textbf{Sequential Embedding and Crossing Check:} The algorithm first embeds the cycle $c$ onto the plane, followed by sequentially embedding each segment. During this process, it checks whether any of the embeddings cross. If crossings occur, $\hat{\mathcal{G}}_{aux}$ is deemed non-planar.
\item \textbf{Handling Non-Planarity:} When $\hat{\mathcal{G}}_{aux}$ is non-planar, the algorithm identifies the embedded segments that intersect with the most recently added segment. It then constructs two planar embeddings: 
\begin{enumerate*}
\item In the first embedding, only the recently added segment is retained, while all conflicting embeddings are removed.
\item In the second embedding, the recently added segment is excluded, preserving all other embeddings.
\end{enumerate*}
\end{itemize}
\noindent A detailed description of the modified Auslander, Parter, and Goldstein algorithm is provided in \ref{append:mod_planar}. Finally, all planar embeddings derived from the non-planar graph $\hat{\mathcal{G}}_{aux}$ are transformed back into planar graphs using Algorithm \ref{alg:planar_nplanar} (see \ref{append:mod_planar}). Let the resulting set of all planar graphs be denoted as $\hat{\mathcal{G}}_{aux}^p$.
\begin{figure*}
	\centering
	\begin{subfigure}[b]{0.2\textwidth}
		\centering
		\includegraphics[width=\textwidth]{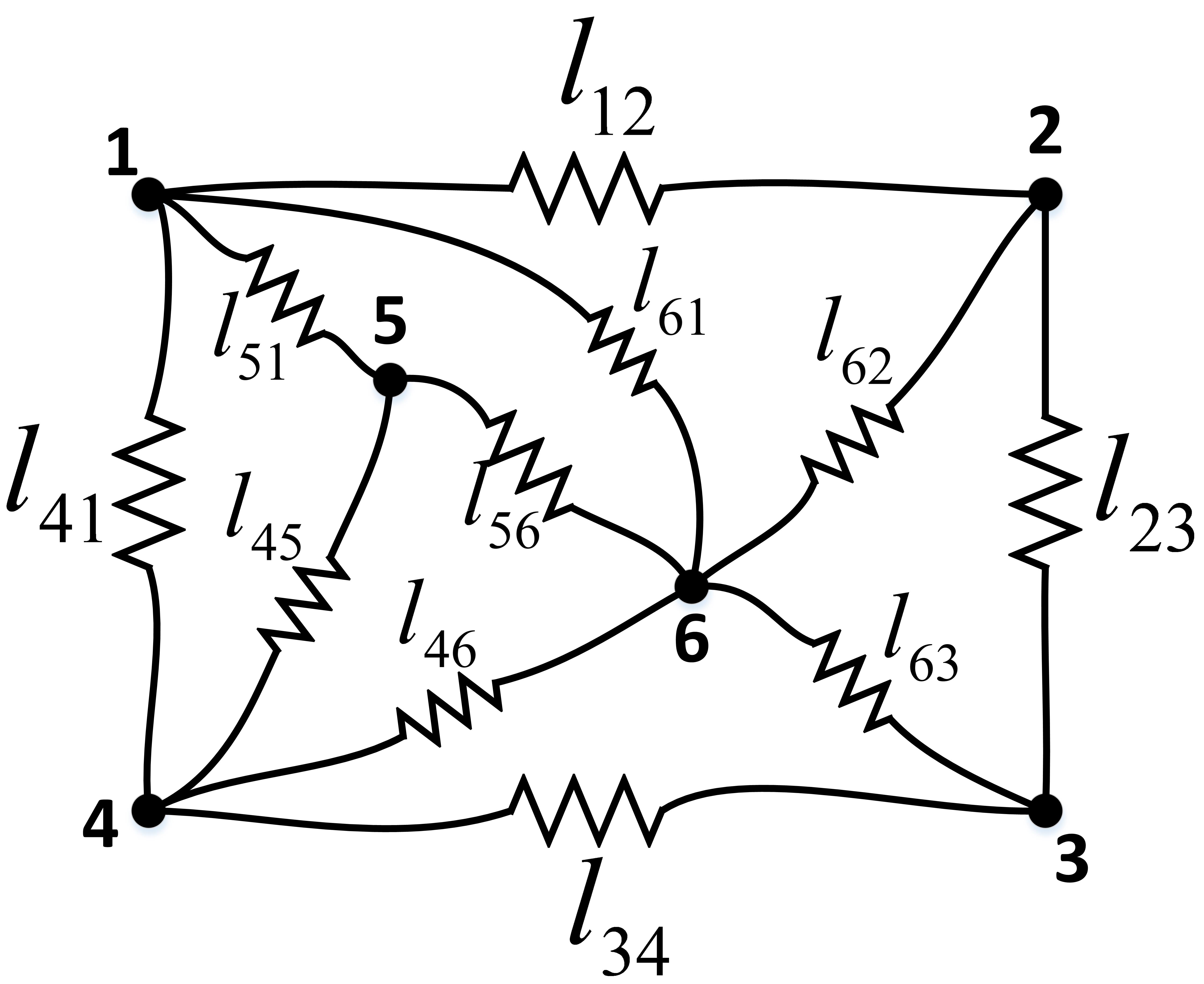}
		\caption{A admissible planar network \texorpdfstring{$\Gamma_1$}{Gamma1}.}
		\label{fig:plane_net_admit1}
	\end{subfigure}
	\hfill
	\begin{subfigure}[b]{0.2\textwidth}
		\centering
		\includegraphics[width=\textwidth]{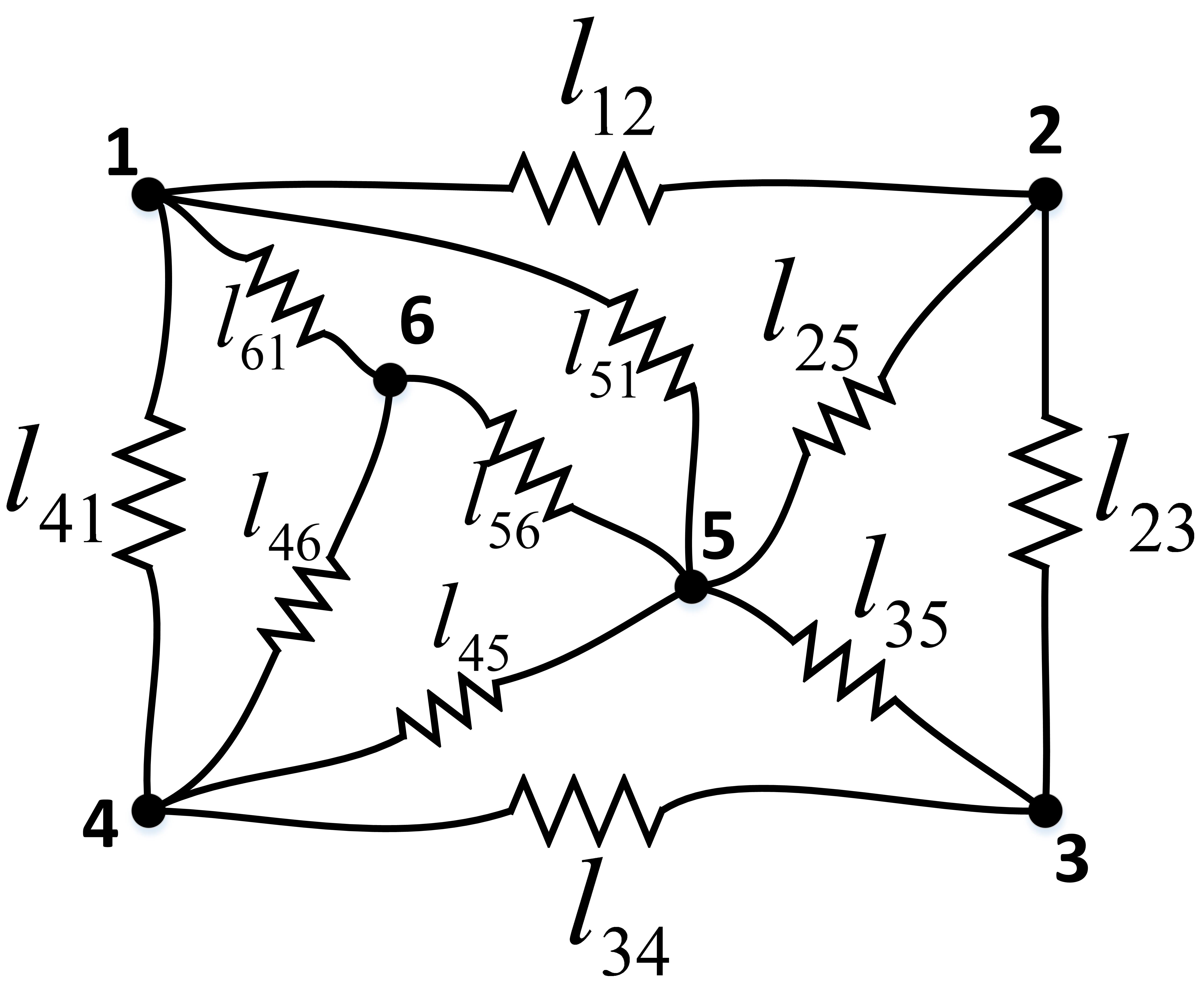}
		\caption{A admissible planar network \texorpdfstring{$\Gamma_2$}{Gamma2}.}
		\label{fig:plane_net_admit2}
	\end{subfigure}
	\hfill
	\begin{subfigure}[b]{0.2\textwidth}
		\centering
		\includegraphics[width=\textwidth]{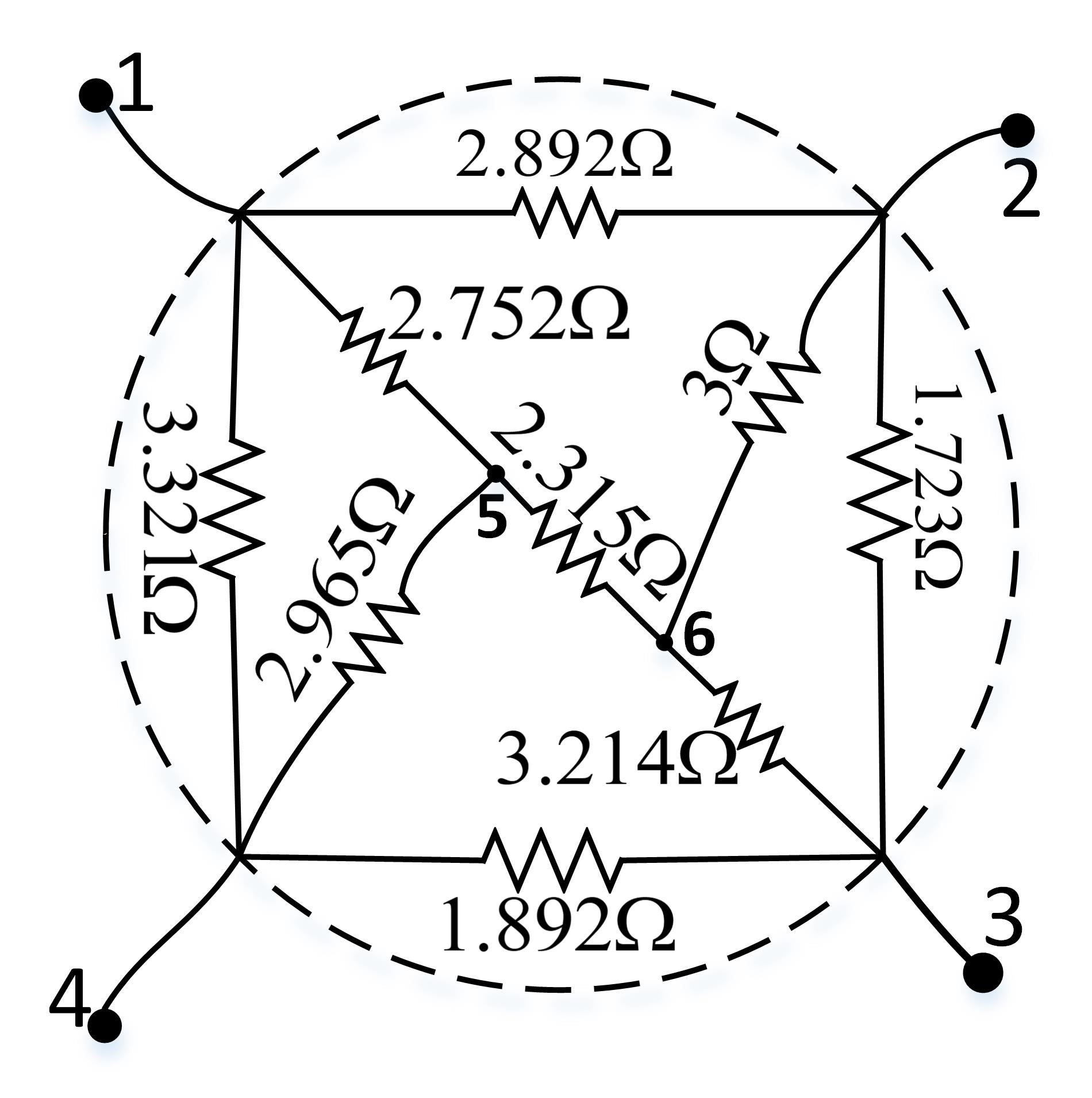}
		\caption{Reconstructed network \texorpdfstring{$\Gamma^*$}{Gamma*}}
		\label{fig:gammastar1}
	\end{subfigure}
	\hfill
	\begin{subfigure}[b]{0.2\textwidth}
		\centering
		\includegraphics[width=\textwidth]{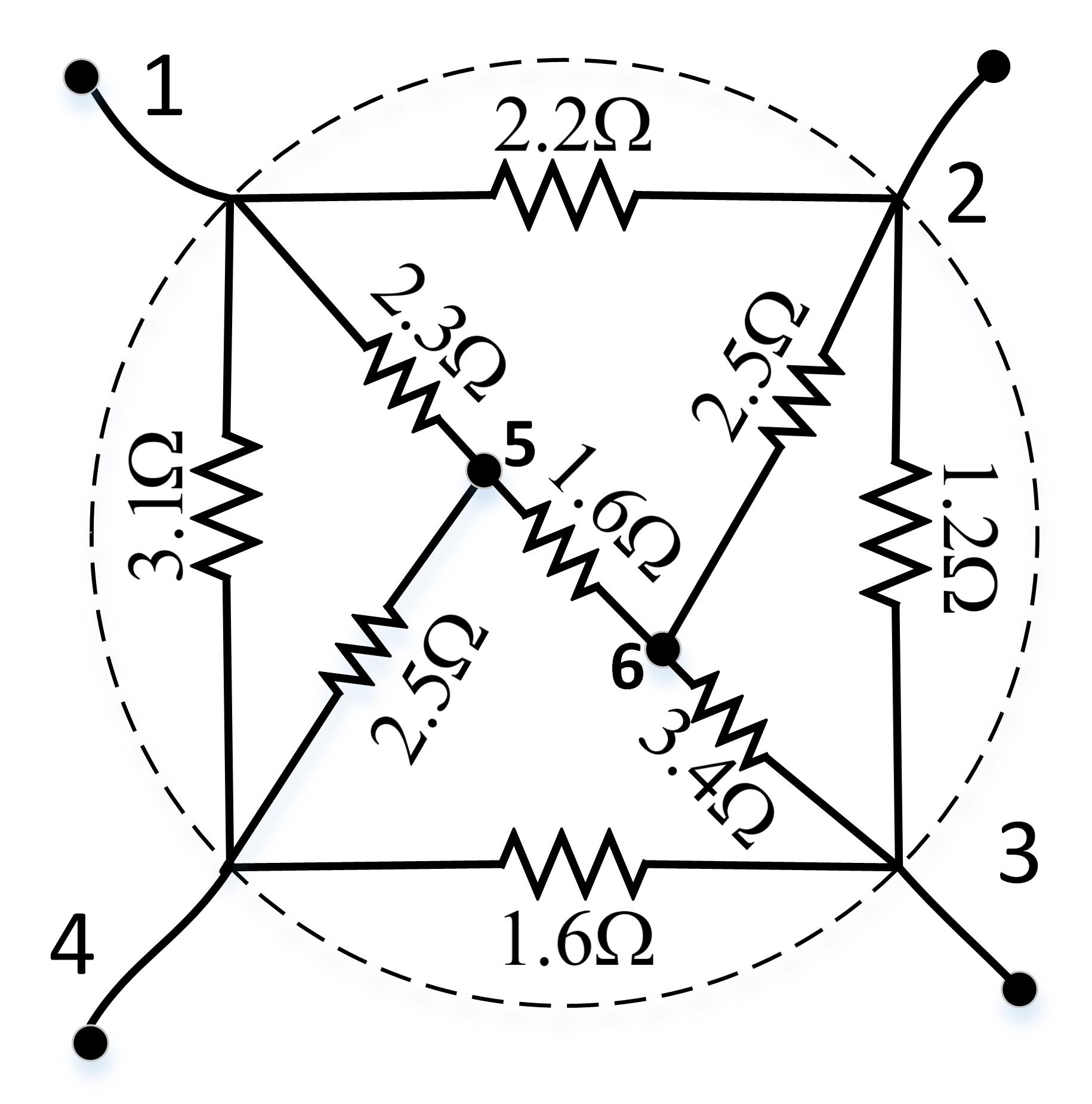}
		\caption{Original \texorpdfstring{$\operatorname{CPPR}$}{CPPR} network.}
		\label{fig:gammastar}
	\end{subfigure}
	\caption{Topology Reconstruction Example.}
	\label{fig:three graphs}
\end{figure*}
\section{Rewiring}\label{rewiring1}
For each graph $\hat{\mathcal{G}}_{aux,i}^{p}$ in the set of planar graphs $\hat{\mathcal{G}}_{aux}^p$, where $\hat{\mathcal{G}}_{aux,i}^{p} = \left( \mathcal{V_B} \cup \mathcal{V_I}, \hat{\mathcal{E}}_{aux,i}^{p} \right)$, we construct a corresponding resistor network $\hat{\Gamma}_i = \left( \hat{\mathcal{G}}_{aux,i}^{p}, \hat{\gamma}_i \right)$. Here, $\hat{\gamma}_i: \hat{\mathcal{E}}_{aux,i}^{p} \to \mathbb{R}^+$ represents the unknown edge conductance function for the $i^{th}$ network, with $1 \leq i \leq |\hat{\mathcal{G}}_{aux}^p|$. Since the conductivity function $\hat{\gamma}_i$ is unknown, we define $\hat{\mathbf{c}}_i$ as the vector of unknown edge conductances for the $i^{th}$ network $\hat{\Gamma}_i$. Our objective is to determine potential rewirings of the network and assign appropriate edge conductances. This is achieved by formulating a sparse difference of convex optimization problem, denoted as $\mathbf{\Pi_3}$, which is presented below.
\begin{equation*}\label{optimiformulationxx}
	\begin{aligned}
	     \min_{\hat{\mathbf{c}}_i,\textbf{W}} \, & ({\tilde{\textbf{r}}^d})^T \textbf{W}  {\tilde{\textbf{r}}^d} + \left( K_{\hat{\Gamma}_{i}}-K_{\Gamma} \right)^2 \hspace{3.2cm} \left(\mathbf{\Pi_3}\right)  \\
		\textrm{s.t} \,\, & \mathbf{0} \preceq \hat{\mathbf{c}}_i \preceq \gamma_{max}\mathbf{1}, \Delta \succeq 0,
		\mathcal{K} \succeq 0,
  0.5 \le W_{ij}\le 0.9\, \forall i,j \in \mathcal{U_B}.  
	\end{aligned}
\end{equation*}
The vector $\hat{\mathbf{c}}_i$ represents the solution to the convex optimization problem $\mathbf{\Pi_3}$. If some elements of $\hat{\mathbf{c}}_i$ fall within the interval $[0, \gamma_{min}]$, we apply a rounding algorithm. This algorithm assigns values in $(0, \gamma_{min})$ to either $0$ or $\gamma_{min}$, based on the sign of the derivative (similar to the approach described in equation (\ref{boolean_derivative})). After modifying $\hat{\mathbf{c}}_i$ accordingly, $\mathbf{\Pi_3}$ is solved again using the updated vector as the initial condition.

For each network $\hat{\Gamma}_i$, we solve $\mathbf{\Pi_3}$ and collect the resulting solution vectors into the set $\hat{\mathbf{c}} = \{\hat{\mathbf{c}}_i : 1 \leq i \leq |\hat{\mathcal{G}}_{aux}^p|\}$. From this set, we select the conductance vector $\mathbf{c}^*$ that minimizes the objective function $({\tilde{\textbf{r}}^d})^T \textbf{W} {\tilde{\textbf{r}}^d} + \left( K_{\hat{\Gamma}_{i}} - K_{\Gamma} \right)^2$. The vector $\mathbf{c}^*$ corresponds to the reconstructed network $\Gamma^*$, which approximates the unknown CPPR network $\Gamma$.
\section{Example}\label{example_prob}
Let us consider an unknown network $\Gamma=\left(\mathcal{G},\gamma\right)$ as shown in Fig.\ref{fig:unk_gamma}. The following are the knowns $n_b=4$, $\mathcal{U}=\{2,5,6\}$, $n_i=2$, $\mathcal{A}=\{1,3,4\}$, $r_{max}=\gamma_{min}^{-1}=4\Omega$, $K_{\Gamma}=19.8\Omega$, and we have $r^d=\{r^d_{1,3}=1.4984\Omega, r^d_{1,4}=1.351\Omega, r^d_{3,4}=1.0795\Omega\}$. We construct an appropriate $MPRSN$ as shown in Fig.\ref{fig:mprsn} in Example \ref{exam_MPRSN}. Formulate an optimization problem $\mathcal{I}$ to compute the estimates $\hat{r}^d_{i,j}\,\forall i,j \in \mathcal{U_B}$ and, then solve $\mathbf{\Pi_1}$ to find an optimal switch combination. The solution to this problem is $\Gamma_{aux}$, shown in Fig.\ref{fig:aux_gamma}. Now, to place interior nodes appropriately, apply heuristic method.
Solution to $\mathbf{\Pi}_2$ is shown in  Fig.\ref{fig:auxbar_gamma}. Then, by examining the solution edge resistance vector $\bar{r}$, interior node $5$ is placed on edge $13$ and interior node $6$ is a dangling node as shown in Fig.\ref{fig:auxhat_gamma}. Now, connect all the interior nodes to every other node to get a network $\hat{\Gamma}$ as shown in Fig.\ref{fig:auxhat_gamma}. The network $\hat{\Gamma}$ may be non planar, we therefore apply modified Auslander, Parter, Goldstein algorithm and construct corresponding planar resistive electrical networks $\hat{\Gamma}_1$ and $\hat{\Gamma}_2$ as shown in Fig. \ref{fig:plane_net_admit1} \& \ref{fig:plane_net_admit2}.
It can be seen that the networks are structurally similar with different numbering for interior nodes. Therefore, solve problem $\mathbf{\Pi}_3$ for $\Gamma_1$ or $\Gamma_2$ to get the solution $\mathbf{c}^*=\hat{\mathbf{c}}_1$ or $\mathbf{c}^*=\hat{\mathbf{c}}_2$. The reconstructed network $\Gamma^{\star}$ corresponding to an unknown $CPPR$ electrical network is shown in Fig. \ref{fig:gammastar1}, and the original network is shown in Fig \ref{fig:gammastar}. The authors would like to acknowledge the availability of the source code related to this example on \href{https://github.com/ShivanB/Topology_Reconstruction_Codes?tab=readme-ov-file#topology_reconstruction_codes}{GitHub}. 
\section{Discussions}
\subsection{Number of Measurements}
In equation (\ref{eqn:laplstru}), if all $|\mathcal{V}| = m$ nodes in 
the network are accessible, the reconstruction problem has a unique 
solution. However, when only boundary nodes are measurable and 
interior nodes remain inaccessible, the system becomes 
underdetermined, leading to infinitely many solutions 
satisfying equation (\ref{eqn:laplstru}). As a result, an 
infinite number of networks can be constructed that align 
with the available resistance distance measurements. To constrain the solution space and ensure physical and geometric 
 validity, we incorporate additional constraints based on the triangle 
 inequality and Kalmanson’s inequality. These constraints help steer 
 the optimization toward solutions that adhere to the expected 
 properties of circular planar resistive networks. However, the 
 proposed multi-stage reconstruction heuristic may not be able to 
 construct all networks satisfying these inequalities, as the algorithm 
 imposes additional structural and computational restrictions at each stage.
\subsection{Methodology}
The proposed methodology is effective for small CPPR networks. However, as the number of boundary nodes increases, the number of edges in the maximal planar graph grows nearly exponentially, with a lower bound on the edge count provided in \cite{tutte1962census}. 

In constructing the maximal planar resistor-switch network (MPRSN), increasing the value of $r_{max}$ leads to a higher number of switch positions that must be optimized, raising the complexity of the optimization problem $\mathbf{\Pi}_1$. To address this, the difference of convex programming (DCCP) approach constructs a global overestimator of the objective function and solves the resulting convex subproblem with relatively low per-iteration computational cost \cite{shen2016disciplined}. The method also relies on simultaneously performing planarity testing and extracting admissible planar embeddings, which becomes computationally demanding for large networks. The number of feasible planar embeddings depends on the distribution of dangling and non-dangling nodes, with the count increasing as the number of dangling nodes rises. The Auslander, Parter, and Goldstein algorithm, which underpins this process, has a computational complexity of $\mathcal{O}(m)$, where $m$ is the number of nodes in the graph.
\subsection{Error}
\begin{figure}
	\centering
	\includegraphics[scale=0.035]{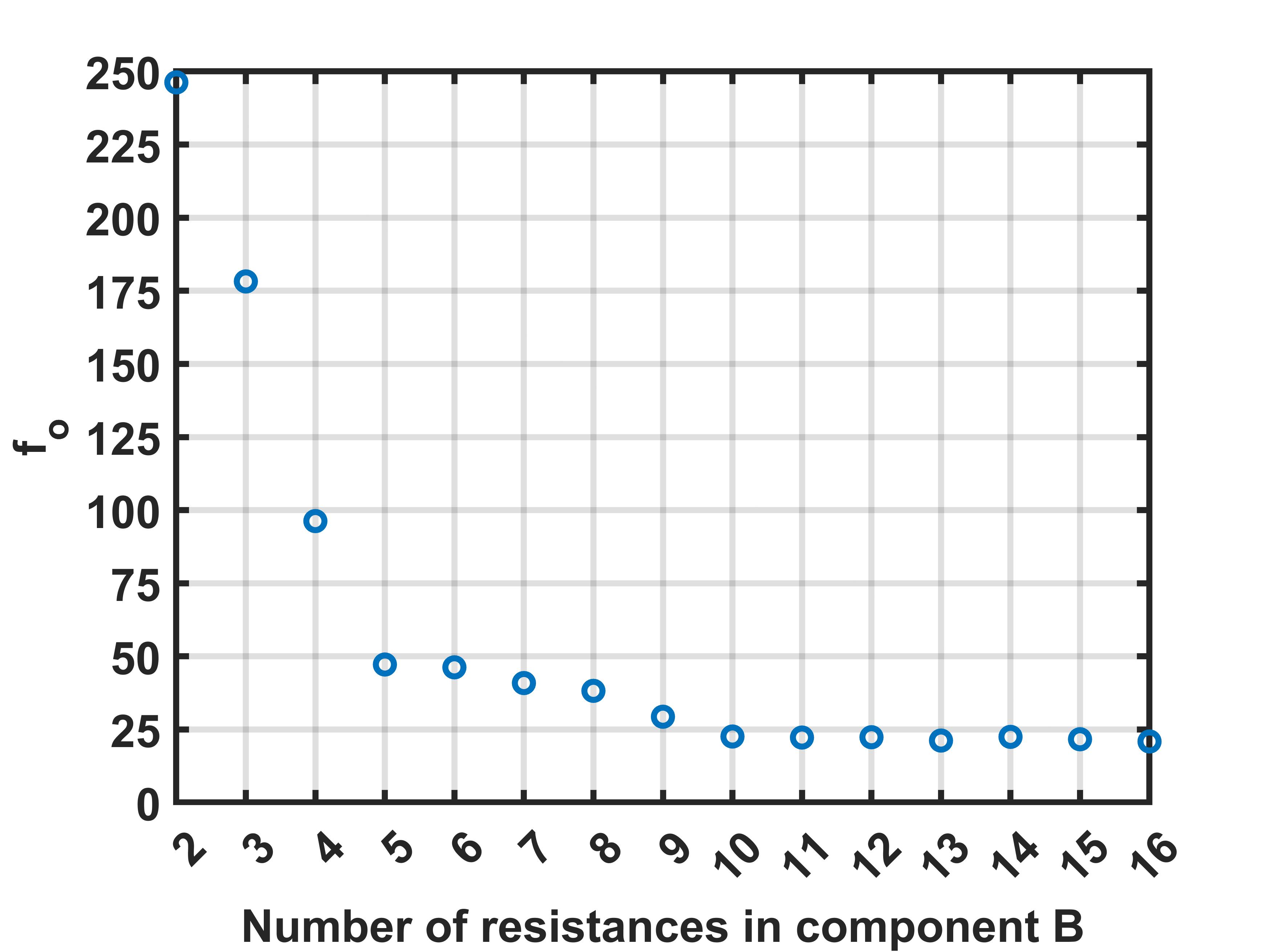}
	\caption{Value of $f_0$ $(\mbox{corresponding to}\hspace{0.1cm} \Gamma^\star)$ w.r.t number of resistances in component B}
	\label{fig:fo}
\end{figure}

Three major sources of errors that induces error in the reconstructed network are, 
\begin{enumerate*}
	\item number of available resistance distance measurements,
	\item number of resistances in component B of the RSN, in $\Gamma_{M}$,
	\item choice of initial conditions for optimization formulation.
\end{enumerate*}
As the number of available resistance distance measurements increases, switch positions can be tuned appropriately to get a better $\Gamma_{aux}$. Therefore, more resistance distance measurements will lead to a reliable $\Gamma_{aux}$, and this $\Gamma_{aux}$ will further lead to proper reconstruction of the network.\\
\indent Let us consider a case wherein the available boundary measurements can properly reconstruct a network. We now consider the effect of the number of resistances in component B of RSN on the value of $f_o$ corresponding to $\Gamma^\star$. It is seen that, as the number of resistances in component B increases, the value of $f_o$ decreases for some values and then remains almost constant, as shown in Fig.\ref{fig:fo}. 
\\ \indent Initial guess values fed into the optimization routine must be chosen judiciously, for proper network reconstruction. The quality of reconstructed network $\Gamma^{\star}$ is sensitive to the choice of guesstimate of switch variables vector $\bm{\rho}$ for $\mathbf{\Pi_1}$. Therefore, a novel method for choosing a proper initial guess of switch positions in $\Gamma_M$ for $\mathbf{\Pi_1}$ is explained in \ref{append:choice_initial}.  
{\subsection{Performance Evaluation Under Noisy Conditions}}
We evaluate the performance of the proposed topology reconstruction method under varying levels of zero-mean additive Gaussian noise. Noise variances are set to 0.125, 0.25, 0.5, 0.75, and 1.0, with 200 independent trials conducted for each level. Performance is measured using the error $||\mathcal{L}_{\mathbf{c}^\star} - \mathcal{L}||$, where $\mathcal{L}_{\mathbf{c}^\star}$ is the Laplacian matrix of the reconstructed network, obtained by solving the optimization problem $\mathbf{\Pi}_3$, and $\mathcal{L}$ is the original Laplacian matrix.
\begin{figure}
    \centering
    \includegraphics[scale=0.02]{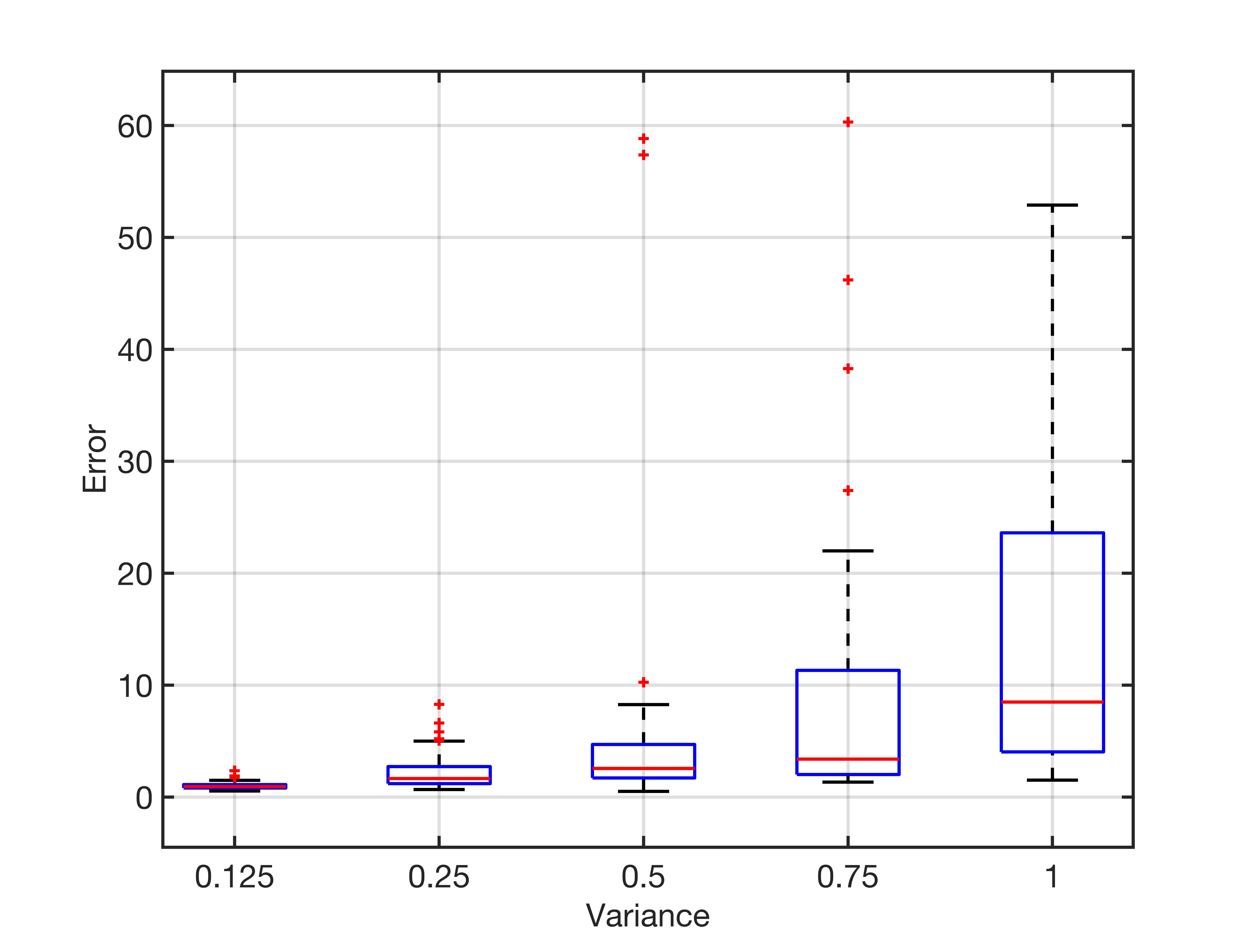}
    \caption{Box Plot Analysis}
    \label{fig:box_plot}
\end{figure}
\noindent Figure \ref{fig:box_plot} presents box plots illustrating the algorithm's performance under different noise levels. At low noise (variance = 0.125), errors are tightly concentrated around zero, with a near-zero median and no significant outliers, indicating robust performance. As noise increases to moderate levels (variances = 0.25, 0.5), the error distribution broadens slightly but remains small, reflecting reasonable resilience. At higher noise levels (variances = 0.75, 1.0), errors become more pronounced, with elevated medians and noticeable outliers. Nonetheless, the algorithm performs acceptably, as most trials stay within an acceptable range. A clear trend of increasing error with rising variance is observed, consistent with expected behavior under noisy conditions. The method demonstrates graceful degradation, maintaining reliable performance up to moderate noise levels.
{\subsection{Convergence and Correctness of Algorithm}}
Let us consider the optimization problem $\mathbf{\Pi_1}$, whose constraints create a compact feasible region $\mathcal{F}$. We express the inequality $\mathcal{K} \ge 0$ w.l.o.g as \(g(\boldsymbol{\rho}) - h(\boldsymbol{\rho})\), where \(g(\boldsymbol{\rho})\) and \(h(\boldsymbol{\rho})\) are convex, and hence it is a \emph{difference of convex} constraint. Therefore, the above optimization problem is a DCCP (Difference of Convex Programming) problem. The DCCP problems are solved using sequential convex programming approach, which decomposes the original non-convex problem into a sequence of convex subproblems. For problem $\mathbf{\Pi}_1$ the convex subproblem is:
\[
\begin{aligned}
\min_{\boldsymbol{\rho}, \mathbf{W}} \quad &f_o\Big(\bm{\rho}, \mathbf{W}\Big)=(\tilde{\mathbf{r}}^d)^T \mathbf{W} \tilde{\mathbf{r}}^d \\
\text{s.t.} \quad & \boldsymbol{\rho} \odot (1-\boldsymbol{\rho}) \geq \mathbf{0}, \quad \Delta \geq 0, 0.5 \leq W_{ij} \leq 0.9,\\
& g(\boldsymbol{\rho}) - \left[ h(\boldsymbol{\rho}^k) + \nabla h(\boldsymbol{\rho}^k)^T (\boldsymbol{\rho} - \boldsymbol{\rho}^k) \right] \geq 0, \\
\end{aligned}
\] 
The sequence of iterates ${\Big(\bm{\rho}^{(k)}, \mathbf{W}^{(k)}\Big)}$ generated by the DCCP algorithm is well-defined and the sequence $\Big\{f_o\Big(\bm{\rho}^{(k)}, \mathbf{W}^{(k)}\Big)\Big\}$ is monotonically non-increasing. Hence, the sequence of iterates $\Big\{\bm\rho^{(k)}, \mathbf{W}^{(k)}\Big\}$ has atleast one accumulation point. 
\begin{prop}\label{prop:limitpoint}(Convergence of Objective Function to a Limit Point)
The function $f_o\Big(\bm{\rho}, \mathbf{W}\Big)$ is convex in $\bm\rho$ and $\mathbf{W}$ independently, and the feasible set $\mathcal{F}$ is a compact. Then, the sequence $\Big\{f_o\Big(\rho^{(k)}, \mathbf{W}^{(k)}\Big)\Big\}$ converges to a limit point $f^\star$.
\end{prop}
Now, we analyze the sequence of iterates $\Big\{\bm{\rho}^{(k)}, \mathbf{W}^{(k)}\Big\}$ and prove that any accumulation point $\Big(\bm{\rho}^*, \mathbf{W}^*\Big)$ is a stationary point of problem $\mathbf{\Pi}_1$. This establishes that the iterates converge to $\Big(\bm{\rho}^*, \mathbf{W}^*\Big)$, satisfying the constraints.
\begin{thm}
Let $\Big(\bm{\rho}^*, \mathbf{W}^*\Big)$ be any accumulation point of the sequence $\Big\{\bm{\rho}^{(k)}, \mathbf{W}^{(k)}\Big\}$ generated by solving sequence of convex problem. If $f_o\Big(\bm{\rho}, \mathbf{W}\Big)$ is convex in a $\bm{\rho}$ and $\mathbf{W}$ independently, the feasible region $\mathcal{F}$ is compact and $f_o\Big(\bm{\rho}, \mathbf{W}\Big)$, $g\Big(\bm{\rho}\Big)$ and $g\Big(\bm{\rho}\Big)$ are Lipschitz continuous then $\Big(\bm{\rho}^*, \mathbf{W}^*\Big)$ is a stationary point of problem $\mathbf{\Pi}_1$.
\end{thm}
\begin{proof}
At each iteration $k_i$, the point $\Big(\bm{\rho}^{(k_i+1)}, \mathbf{W}^{(k_i+1)}\Big)$ satisfies the KKT conditions for the convex subproblem $\Pi_1^{(k_i)}$. Specifically, there exist Lagrange multipliers $\lambda_1^{(k_i+1)} \geq 0$, $\lambda_2^{(k_i+1)} \geq 0$, $\mu^{(k_i+1)} \geq 0$, and $\nu^{(k_i+1)} \geq 0$ such that:
\[
   \nabla_{\bm\rho} f_o\Big(\bm\rho^{(k_i+1)}, \mathbf{W}^{(k_i+1)}\Big) - \lambda_1^{(k_i+1)} \nabla c_1\Big(\bm\rho^{(k_i+1)}\Big) 
\]
\[  \lambda_2^{(k_i+1)} \nabla c_2\Big(\bm\rho^{(k_i+1)}\Big) -\sum_{j=1}^J \mu_j^{(k_i+1)} \Big[\nabla g_j(\bm\rho^{(k_i+1)}) - \nabla h_j(\bm\rho^{(k_i)})\Big] = 0,\]
where $c_1=\bm\rho \odot \Big(1-\bm\rho\Big)$, constraint $c_2=\Delta$ \Big($\Delta$ is triangle inequality \Big). For unknown variable $W$, following is true,
\[
   \nabla_\mathbf{W} f_o\Big(\bm\rho^{(k_i+1)}, \mathbf{W}^{(k_i+1)}\Big) - \nu^{(k_i+1)} \nabla c_3\Big(\mathbf{W}^{(k_i+1)}\Big) = 0.
\]
The complementary slackness conditions are:
\[
\lambda_1^{(k_i+1)} c_1\Big(\bm\rho^{(k_i+1)}\Big) = 0, \quad \lambda_2^{(k_i+1)} c_2\Big(\bm\rho^{(k_i+1)}\Big) = 0,
\]
\[
\mu_j^{(k_i+1)} \left[g_j\Big(\bm\rho^{(k_i+1)}\Big)\right] - h_j\Big(\bm\rho^{(k_i+1)}\Big)- 
\]
\[
\nabla h_j\Big(\bm\rho^{(k_i+1)}\Big)^T \Big(\bm\rho^{(k_i+1)} - \bm\rho^{(k_i+1)}\Big) = 0,
\]
\[
\nu^{(k_i+1)} c_3\Big(\mathbf{W}^{(k_i+1)}\Big) = 0.
\]
The gradients of the constraints and the objective function are Lipschitz continuous, and the feasible set $\mathcal{F}$ is compact. Thus, the sequences of multipliers $\Big\{\lambda_1^{(k_i+1)}\Big\}$, $\Big\{\lambda_2^{(k_i+1)}\Big\}$, $\Big\{\mu_j^{(k_i+1)}\Big\}$, and $\Big\{\nu^{(k_i+1)}\Big\}$ are bounded. Consequently, there exists a subsequence such that:
\[
\lim_{i \to \infty} \lambda_1^{(k_i+1)} = \lambda_1^*, \quad \lim_{i \to \infty} \lambda_2^{(k_i+1)} = \lambda_2^*, \quad
\lim_{i \to \infty} \mu_j^{(k_i+1)} = \mu_j^*, 
\]
\[\quad \lim_{i \to \infty} \nu^{(k_i+1)} = \nu^*.\]
Taking the limit as $i \to \infty$ in the KKT conditions for $\mathbf{\Pi}_1^{(k_i)}$, and using the continuity of the gradients of $f_o$ and the constraint functions, we obtain:
\begin{itemize}
    \item[$\bullet$] Gradient Condition for $\bm\rho$:\[
   \nabla_{\bm\rho} f_o\Big({\bm\rho}^*, \mathbf{W}^*\Big) - \lambda_1^* \nabla c_1\Big({\bm\rho}^*\Big) - \lambda_2^* \nabla c_2\Big({\bm\rho}^*\Big) 
   - \]
   \[\sum_{j=1}^J \mu_j^* \left[\nabla g_j\Big({\bm\rho}^*\Big) - \nabla h_j\Big({\bm\rho}^*\Big)\right] = 0,\]
   \item[$\bullet$] Gradient Condition for $\mathbf{W}$:
      \[
   \nabla_\mathbf{W} f_o\Big({\bm\rho}^*, \mathbf{W}^*\Big) - \nu^* \nabla c_3\Big(\mathbf{W}^*\Big) = 0,
   \]
   \item[$\bullet$] Complementary Slackness Conditions:
   \[
   \lambda_1^* c_1\Big({\bm\rho}^*\Big) = 0, \quad \lambda_2^* c_2\Big({\bm\rho}^*\Big) = 0,
   \]
   \[
   \mu_j^* \left[g_j\Big({\bm\rho}^*\Big) - h_j\Big({\bm\rho}^*\Big)\right] = 0, \quad \nu^* c_3\Big(\mathbf{W}^*\Big) = 0.
   \]
\end{itemize}
These are precisely the KKT conditions for the original problem $\mathbf{\Pi}_1$ at the point $\Big({\bm\rho}^*, \mathbf{W}^*\Big)$. Since $\Big({\bm\rho}^*, \mathbf{W}^*\Big)$ satisfies the KKT conditions of $\mathbf{\Pi}_1$, it is a stationary point of the problem. This completes the convergence and correctness proof. Same analysis can be extended for optimization problem $\mathbf{\Pi}_2$ and $\mathbf{\Pi}_3$.
\end{proof}
{\subsection{Unbiasedness of Algorithm}}
\noindent Consider the optimization problem is $\mathbf{\Pi}_1$, wherein the objective function $f_o\Big(\bm{\rho}, \mathbf{W}\Big)=(\tilde{\mathbf{r}}^d)^T \mathbf{W} \tilde{\mathbf{r}}^d$, here \( \bm{\tilde{r}^d} = \bm{r^d} - \bm{\hat{r}^d(\rho)} \). Let $\bm{{r}^{d,true}}$ be the vector of true measurement and $\mathbf{r^d}$ be the vector of noisy measurements, defined as 
$\bm{r^d} = \bm{r^{d,\text{true}}} + \bm{\epsilon},$
where \( \bm{\epsilon} \sim \mathcal{N}(\bm{0}, \bm{\Sigma}) \) is a zero-mean Gaussian noise with covariance \( \bm{\Sigma} \). 
Consider the objective function $f_o\Big(\bm{\rho}, \mathbf{W}\Big)$, which is rewritten as:
\[
f_o\Big(\bm{\rho}, \mathbf{W}\Big) = \Big( \bm{r^{d,true}} + \bm{\epsilon} - \bm{\hat{r}^d(\rho)} \Big)^T \bm{W} \Big( \bm{r^{d,true}} + \bm{\epsilon} - \bm{\hat{r}^d(\rho)} \Big)
\]
\noindent Expanding the objective function \( f_o\Big(\bm{\rho}, \mathbf{W}\Big) \), we get:
\[
f_o\Big(\bm{\rho}, \mathbf{W}\Big) = {\Big( \bm{r^{d,true}} - \bm{\hat{r}^d(\rho)} \Big)^T \bm{W} \Big( \bm{r^{d,true}} - \bm{\hat{r}^d(\rho)} \Big)} + 
\]
\[{2\bm{\epsilon}^T \bm{W} \Big( \bm{r^{d,true}} - \bm{\hat{r}^d(\rho)} \Big)} + {\bm{\epsilon}^T \bm{W} \bm{\epsilon}}.\]
\noindent Now, take the expectation of the objective function,
\[
\mathbb{E}\Big[f_o\Big(\bm{\rho}, \mathbf{W}\Big)\Big] = \Big( \bm{r^{d,true}} - \bm{\hat{r}^d(\rho)} \Big)^T \bm{W} \Big( \bm{r^{d,true}} - \bm{\hat{r}^d(\rho)} \Big) + 
\]
\[\mathbb{E}[\bm{\epsilon^T W \epsilon}].\]
The cross term \( \mathbb{E}\Big[{2\bm{\epsilon}^T \bm{W} \Big( \bm{r^{d,true}} - \bm{\hat{r}^d(\rho)} \Big)}\Big] = 0 \), since \( \mathbb{E}[\bm\epsilon] = 0 \), and 
the covariance term \( \mathbb{E}\Big[\bm{\epsilon}^T \bm{W} \bm{\epsilon}\Big] = \text{Tr}\Big(\bm{W \Sigma}\Big) \), is independent of \( \bm\rho \). Thus, minimizing \( \mathbb{E}\Big[f_o\Big(\bm{\rho}, \mathbf{W}\Big)\Big] \) for $\bm{\rho}$ is equivalent to minimizing: $\Big( \bm{r^{d,true}} - \bm{\hat{r}^d(\rho)} \Big)^T \bm{W} \Big( \bm{r^{d,true}} - \bm{\hat{r}^d(\rho)} \Big).$
Therefore, this is minimized when \( \bm{\hat{r}^d(\rho) = r^{d,true}} \), which occurs at \( \bm{\rho = \rho^{true}} \). While, minimizing for the $\mathbf{W}$ is equivalent to minimizing $\mathbb{E}\Big[\bm{\epsilon}^T \bm{W} \bm{\epsilon}\Big].$

\noindent The constraints in $\mathbf{\Pi}_1$ are derived from physical properties of resistance networks. Since \(\bm{\rho_{true}}\) satisfies these constraints, it lies within the feasible set. For zero-mean noise, the weighted noise term \( \bm{\epsilon^T W \epsilon} \) does not introduce bias, as \( \mathbb{E}\Big[\bm{\epsilon^T W \epsilon}\Big] \) depends only on \( \textbf{W} \) and \( \bm\Sigma \), not \( \bm\rho \). As the noise variance \( \bm\Sigma \to 0 \) and/or the number of measurements increases: 1) The noise term \( \bm\epsilon \) becomes negligible, the solution \( \bm{\hat{\rho}} \to \bm{\rho_{true}} \), ensuring unbiasedness. This proves the unbiasedness of the estimator.
{\subsection{Sensitivity Analysis}}
 Let $\bm{\Delta r^d} = \bm{\epsilon}$ represent the perturbation in the measurement due to noise. The change in the objective function $\Delta f_o(\bm\rho, \mathbf{W}) = \nabla_{\bm{r^{d}}}f_o(\bm{\rho}, \mathbf{W}) \Delta \mathbf{r^d}$. Here, $$\nabla_{\bm{r^{d}}}f_o(\bm{\rho}, \mathbf{W}) = 2 \mathbf{W} \Big(\bm{r^{d}} - \bm{\hat{r}^{d}(\rho)}\Big).$$
    Therefore,
    $$\nabla_{\bm{r^{d}}}f_o(\bm{\rho}, \mathbf{W}) = 2 \mathbf{W} \Big(\bm{r^{d,true}} - \bm{\hat{r}^{d}(\rho)} + \bm{\epsilon}\Big) \bm{\Delta r^d}.$$
The first order sensitivity of $f_o(\bm\rho, \mathbf{W})$ is directly proportional to $\mathbf{W}\bm{\epsilon}$. Whereas, the sensitivity of gradient 
$\nabla_{\bm\rho} f_o(\bm\rho, \mathbf{W})$ with respect to $\bm\epsilon$ is,
$$\nabla_{\bm\rho} f_o(\bm\rho, \mathbf{W}) = -2\mathbf{W}\Big( \bm{r^{d,true} - \bm{\hat{r}^{d}(\rho)}} \Big) \mathbf{W}\,\nabla_{\rho}\bm{r^d(\bm\rho)} - $$ 
\[2\mathbf{W}\bm\epsilon\, \nabla_{\bm\rho} \Big( \bm{r^d(\rho)} \Big)\]
The term $-2\mathbf{W}\Big( \bm{r^{d,true}} - \bm{\hat{r}^{d}(\rho)} \Big) \mathbf{W}\nabla_{\rho}\bm{r^d}(\bm\rho)$ captures the sensitivity of the objective function to the deterministic error between true and reconstructed resistance distances, while the term $- 2\mathbf{W}\bm\epsilon \nabla_{\bm\rho} \Big( \bm{\hat{r}^{d}(\rho)} \Big)$ reflects its sensitivity to measurement noise $\bm\epsilon$.
Finally, the sensitivity of gradient $\nabla_{\bm{W}} f_o(\bm\rho, \mathbf{W})$ with respect to $\bm{W}$ is,
\[
\nabla_{\bm{W}} f_o(\bm\rho, \bm{W}) = \Big(\bm{r^{d,true}} - \bm{\hat{r}^{d}(\rho)}\Big)\Big(\bm{r^{d,true}} - \bm{\hat{r}^{d}(\rho)}\Big)^T + 
\]
\[2 \Big(\bm{r^{d,true}} - \bm{\hat{r}^{d}(\rho)}\Big) \bm{\epsilon}^T + \bm{\epsilon} \bm{\epsilon}^T.\]
\noindent The sensitivity of $\nabla_{\bm{W}} f_o(\bm\rho, \bm{W})$ to noise depends on the magnitude of $\bm{\epsilon}$ and its interaction with $\Big(\bm{r^{d,true}} - \bm{\hat{r}^{d}(\rho)}\Big)$.
\subsection{Initial Conditions}
The problems $\mathbf{\Pi}_1$, $\mathbf{\Pi}_2$ \& $\mathbf{\Pi}_3$ are implemented using a $DCCP$ and is sensitive to initial guess. Therefore, a proper choice of initial guess is necessary to get the right solution. In problem $\mathbf{\Pi}_1$, the initial $\bm{\rho}^{(0)} \in \{0,1\}^p$ is chosen by applying algorithms mentioned in \ref{append:choice_initial}. The solution to the problem $\mathbf{\Pi}_1$ is $\Gamma_{aux}$. For problem $\mathbf{\Pi}_2$, the initial guesstimate $\mathbf{\bar{c}}^{(0)} \in \mathbf{R}^{|\mathcal{E}_{aux}| \times 1}$ is based on the edge resistances of $\Gamma_{aux}$. Therefore, the $l^{th}$ element of $\mathbf{\bar{c}}^{(0)}$, which is also the $l^{th}$ edge of $\Gamma_{aux}$ is  $\mathbf{\bar{c}}_{l}^{(0)}=\gamma_{aux}(l)$. For $\mathbf{\Pi}_3$, the initial guesstimate $\mathbf{\hat{c}}_i^{(0)} \in \mathbf{R}^{|\mathcal{\hat{E}}| \times 1}$ is inferred from the edge resistances of network $\bar{\Gamma}_{aux}$ and $\gamma_{max}$. The $\mathbf{\hat{c}}_i^{(0)}$ is decided as follows, the edge conductance $\mathbf{\hat{c}}^{(0)}_{i}(\sigma)=\Bar{\gamma}(l)$ if $\sigma \in \mathcal{E}_{aux}$, $\mathbf{\hat{c}}^{(0)}_{i}(\sigma)={\gamma}_{max}$ if $\sigma \in \mathcal{E}_{p}$ and $\mathbf{\hat{c}}^{(0)}_{i}(\sigma)=0$ if $\sigma \notin \mathcal{E}_{aux}$. 
\section{Conclusion} 
We propose a multistage algorithm to reconstruct the topology of a general CPPR 
network. The algorithm assumes limited information: only some resistance distance 
measurements, the number of boundary and interior nodes, the minimum and maximum 
edge conductance, and the Kirchhoff index are known. The process begins by 
constructing an initial network, $\Gamma_{aux}$, in two steps: first, creating a 
maximal planar network with edges containing resistors and switches configured 
based on the maximum resistance value; second, determining switch positions using 
available resistance distance measurements. This is achieved by solving a difference 
of convex programming problem, $\mathbf{\Pi}_{1}$, which incorporates a quadratic 
cost function constrained by the triangle inequality and Kalmanson’s inequalities. 
The resulting $\Gamma_{aux}$ serves as the foundation for subsequent stages. Since 
$\Gamma_{aux}$ does not include interior nodes, we optimize its edge resistances by 
solving another optimization problem, $\mathbf{\Pi}_{2}$, which refines 
$\mathbf{\Pi}_{1}$ with additional constraints on the Kirchhoff index and 
relaxed conductance limits. Based on the optimized resistances, interior nodes are
 added to edges exceeding the maximum allowable resistance, while remaining interior 
 nodes are treated as dangling nodes. The connections among interior nodes and between 
 interior and boundary nodes are initially unknown. To address this, all interior nodes
  are connected to every other node, potentially resulting in a non-planar network. We 
  then apply a modified Auslander, Parter, and Goldstein algorithm to extract planar
   networks. For each planar network, edge conductances are computed by solving an 
   optimization problem similar to $\mathbf{\Pi}_{1}$. The network that best matches 
   the available resistance distance measurements and the Kirchhoff index is selected
    as the reconstructed network. From a control-design perspective, the reconstructed model facilitates 
	model-based control synthesis, optimal actuator and sensor placement 
	informed by the revealed topology, and the design of fault-tolerant 
	systems through sequential monitoring. This methodology is particularly effective for 
	networks with fewer nodes. However, aspects like the computation of 
	$\Gamma_{aux}$, placement of interior nodes, and planar network construction 
	can be improved for greater efficiency. Additionally, while the Kirchhoff 
	index is assumed known here, it is often unavailable in practice. Bounds on 
	the Kirchhoff index \cite{bianchi2013bounds} can be incorporated into the 
	optimization to address this limitation. {Finally, the approach can be extended 
	to reconstruct RLC networks, and general networks offering a direction for future 
	research.}
\bibliography{autosam}                

\appendix
\section{Construction of Initial Guess for $\mathbf{\Pi_1}$}\label{append:choice_initial}
The initial guess fed into $\mathbf{\Pi_1}$ are the initial switch positions in $\Gamma_M$. We present a novel algorithm to compute an initial guess $\bm{\rho}^{(0)} \in \{0,1\}^{t}$, where $t=\left(3n_b-6\right)\left(\left \lfloor{r_{max}}\right \rfloor - 1\right) + 10$. The algorithm comprises of solving $\mathcal{I}$ first. Then, using $\textbf{r}^d$ and the estimated resistance distances $\hat{r}^d_{i,j} \forall i,j \in \mathcal{U_B}$, an iterative algorithm is run to compute $\bm{\rho}^{(0)}$, i.e., the initial switch positions. \\
The estimated resistance distances $\hat{r}^d_{ij}=\hat{\textbf{R}}_{\Gamma}\left(i,j\right), \forall i,j \in \mathcal{U_B}$ computed from $\mathcal{I}$ and the available resistance distances in set $\textbf{r}^d$ are used in the proposed iterative algorithm to get initial guess vector $\bm{\rho}^{(0)}$.
The iterative algorithm involves,
\begin{enumerate*}
	\item[1.)] computation of edge resistances, by increasing and decreasing edge resistance by $1\Omega$,  
	\item[2.)] addition and deletion of edges,
\end{enumerate*}
based on $\textbf{r}^d$ and $\hat{r}^d_{i,j} \forall i,j \in \mathcal{U_B}$. The algorithm is designed specifically to assign only integer edge resistances upto value $r_{max}$. The iterative algorithm gives an electrical network from which an initial switch position guess $\bm{\rho}^{(0)}$ is determined to be fed into $\mathbf{\Pi}_1$.\\  
\indent At $0^{th}$ iteration, we consider network $\Gamma_{I}^{(0)}=\left(\mathcal{G}_I^{(0)},\gamma^{(0)} \right)$, where $\mathcal{G}_I^{(0)}=\mathcal{G}_{n_b}^{max}$. Instead of using edge conductance, we use edge resistance for explanation in this section. The edge resistances are set to $r^{(0)}\left(ij\right)=1\Omega,\, \forall ij \in \mathcal{E}^{max}$. Let the corresponding Laplacian matrix be $\mathcal{L}\left(\Gamma_I^{(0)}\right)$. Then, set of resistance distances corresponding to $\Gamma_I^{(0)}$ is $r^{d}\left(\Gamma_I^{(0)}\right)\triangleq \left\{r^{d}_{i,k}\left(\Gamma_I^{{(0)}}\right)=\textbf{b}_{ik}^T \mathcal{L}\left(\Gamma_I^{(0)}\right)^{\dagger}\textbf{b}_{ik}| i,k \in \mathcal{V_B}\right\}$, and the resistance distance error set is
$${\tilde r^{d,\left( 0 \right)}} = \left\{ \begin{array}{l}
	\tilde r_{i,k}^{d,\left( 0 \right)} =  r_{i,k}^d\left( {\Gamma _I^{\left( 0 \right)}} \right) - r_{i,k}^d{\rm{, if }}\,i,k \in \mathcal{A}\,{\rm{ or}}\\
	{\rm{ }}\tilde r_{i,k}^{d,\left( 0 \right)} =  r_{i,k}^d\left( {\Gamma _I^{\left( 0 \right)}} \right) - \hat r_{i,k}^d,{\rm{ if }}\,i,k \in \mathcal{U_B} {\rm{ }}
\end{array} \right\}$$.  Since algorithm involves deletion and addition of edges we keep track of added and deleted edges, using $\mathcal{D}^{(0)}$, the set of deleted edges and $A^{(0)}$ be the set of edges added at $0^{th}$ iteration. Initially, $\mathcal{D}^{(0)}=\emptyset$ and $A^{(0)}=\emptyset$. Let $d_i^{(0)}$ be the degree of node $i$ at $0^{th}$ iteration. If degree of any node in an edge is $1$ we call such edge as a floating edge. The algorithm starts with identifying nodes pairs, say $s,t \in \mathcal{V_B}$, across which the maximum absolute resistance distance error occurs and $st \in \mathcal{E}^{max}$ ($\mathcal{E}^{max}$ is defined in section \ref{sec:initalnetwork}). Now, the aim is to increase or decrease the edge resistance $r^{(0)}(st)$ such that $\tilde{r}^{d,(0)}_{s,t}$ is minimized. Hence, if $\tilde{r}^{d,(0)}_{s,t}<0$, then we either delete the edge $st$ or increase the edge resistance $r^{(0)}(st)$ by $1\Omega$, whichever is better. Whereas, if $\tilde{r}^{d,(0)}_{s,t}>0$ then we either add an edge with edge resistance $1\Omega$ or decrease edge resistance value by $1\Omega$. This is exemplified  for $n^{th}$ iteration, given below.
\\           
\indent At $n^{th}$ iteration, consider a network $\Gamma_I^{(n)}=\left(\mathcal{G}_I^{(n)}, \gamma^{(n)}\right)$, where $\mathcal{G}_I^{(n)}=\left(\mathcal{V_B},\mathcal{E}_I^{(n)}\right)$ and $r^{(n)}:\mathcal{E}_I^{(n)} \rightarrow \mathbf{Z}^+_{\le r_{max}}$. Then, ${r^d}\left( {\Gamma _I^{\left( n \right)}} \right) = \left\{ \begin{array}{l}
	r_{i,k}^d\left( {\Gamma _I^{\left( n \right)}} \right) = \textbf{b}_{ik}^T\mathcal{L}{\left( {\Gamma _I^{\left( n \right)}} \right)^ \dagger }{\textbf{b}_{ik}}:\\
	i,k \in {\mathcal{V_B}}
\end{array} \right\}$ is a set of resistance distances corresponding to $\Gamma_I^{(n)}$, and a resistance distance error set
${\tilde r^{d,\left( n \right)}} = \left\{ \begin{array}{l}
	\tilde r_{i,k}^{d,\left( n \right)} = r_{i,k}^d\left( {\Gamma _I^{\left( n \right)}} \right) - r_{i,k}^d{\rm{, if }}\,i,k \in \mathcal{A}\,{\rm{ or}}\\
	{\rm{ }} \tilde r_{i,k}^{d,\left( {n} \right)} = r_{i,k}^d\left( {\Gamma _I^{\left( n \right)}} \right) - \hat r_{i,k}^d,{\rm{ if }}\,i,k \in \mathcal{U_B}{\rm{ }}\end{array} \right\}$. Also, let $\mathcal{D}^{(n)}$ and $A^{(n)}$ be non empty set, then $\mathcal{E}_I^{(n)}=\left(\mathcal{E}^{(0)} \setminus \mathcal{D}^{(n)}\right) \bigcup A^{(n)}$. Now choose an index pair, say $\{s,t\} \in \mathcal{V_B}$, across which maximum absolute resistance distance error occurs from set $\tilde{r}^{d,(n)}$, let us denote this process of choosing $\{s,t\}$  as, $\{s,t\}=\textbf{indexmax}\left( \tilde{r}^{d,(n)} \right) $. Then, based on the sign of $\tilde{r}^{d,(n)}_{s,t}$ and various other criteria, several operations on  graph $\mathcal{G}^{(n)}_{I}$ are executed at $n^{th}$ iteration. That is, if \begin{enumerate*}
 \item[1.)] if $\tilde{r}^{d,(n)}_{s,t}<0$ and $st \in \mathcal{E}_I^{(n)}$, then we either delete an edges $st$ (let us call this operation $\mathcal{OP}1$)  or increase the edge resistance by $1\Omega$ (let us call this operation $\mathcal{OP}2$). Both operations are given in details as Algorithm-\ref{algo_initialguess1} and \ref{algo_initialguess2} respectively. 
 \item[2.)] if $\tilde{r}^{d,(n)}_{s,t}<0$ and $st \notin \mathcal{E}_I^{(n)}$ then, find another node pair, say $\{s_1,t_1\}$, such that $\{s_1,t_1\}=\textbf{indexmax}\left( \tilde{r}^{d,(n)} \right) $ and $s_1t_1 \ne st$. 
 \item[3.)] if $\tilde{r}^{d,(n)}_{s,t}>0 \land st \notin \mathcal{E}_I^{(n)}$ then, we add a new edge $st$ across nodes $s$ and $t$. This operation is called as $\mathcal{OP}3$ and is presented in details as Algorithm-\ref{algo_initialguess3}.
 \item [4.)] if $\tilde{r}^{d,(n)}_{s,t}>0 \land st \in \mathcal{E}_I^{(n)}$ then we decrease the edge resistance $r^{(n)}(st)$ by $1\Omega$. Let this operation be named as $\mathcal{OP}4$ and is given in detail as Algorithm-\ref{algo_initialguess4}.
 \end{enumerate*}\\
 Each operation is briefly explained case by case below,\\
\textbf{Case-1:}\,if $\tilde{r}^{d,(n)}_{s,t}<0$ and $st \in \mathcal{E}_I^{(n)}$ then,\\  \begin{enumerate*} \item[1.] the edge $st$ can be deleted, or, \item[2.] edge resistance $r^{(n)}(st)$ is increased by $1\Omega$.\end{enumerate*} Let us call an operation in point $1$ as $\mathcal{OP}1$ and, operation in point $2$ as $\mathcal{OP}2$. In case-1, first implement operation $\mathcal{OP}1$, then operation $\mathcal{OP}2$ on $\Gamma_I^{(n)}=\left(\mathcal{G}_I^{(n)}, \gamma^{(n)}\right)$ independently. Let $\tilde{r}^d_{s,t}\left(\mathcal{OP}1\right)$ and $\tilde{r}^d_{s,t}\left(\mathcal{OP}2\right)$ be the resistance distance error across $s,t \in \mathcal{V_B}$ after committing operation $\mathcal{OP}1$ and $\mathcal{OP}2$ respectively.\\ 
\indent An operation is said to be valid if a committed operation results in improvement of resistance distance error, i.e., if $\left|\tilde{r}^d_{s,t}\left(\mathcal{OP}1\right)\right|<\left|\tilde{r}^{d,(n)}_{s,t}\right|$ then $\mathcal{OP}1$ is a valid operation to commit. Now, if both $\mathcal{OP}1 \, \mbox{and} \,  \mathcal{OP}2$ are valid, then choose an operation which results in $\min\left\{|\tilde{r}^d_{s,t}\left(\mathcal{OP}1\right)|, |\tilde{r}^d_{s,t}\left(\mathcal{OP}2\right)|\right\}$. If only any one of the operation is valid, then implement that operation. If both are invalid then find another node pair, say $s_1,t_1$, such that $\left\lbrace s_1,t_1 \right\rbrace=\textbf{indexmax}\left( \tilde{r}^{d,(n)} \right)$ and $s_1t_1 \notin \mathcal{D}^{(n)}$. Therefore, let us define a function $\textbf{OPselect}\left(\mathcal{OP}1, \mathcal{OP}2\right)$, which helps select an appropriate operation based on $\tilde{r}^d_{s,t}\left(\mathcal{OP}1\right)$, $\tilde{r}^d_{s,t}\left(\mathcal{OP}2\right)$ and $\tilde{r}^{d,{(n)}}_{s,t}$ as explained above. This function is used in Algorithm-\ref{algo_initialguess2} and is explained in Algorithm-\ref{algo_initialguess5}.
The operations $\mathcal{OP}1$ and $\mathcal{OP}2$ are given in details as Algorithm-\ref{algo_initialguess1} and \ref{algo_initialguess2}.
\begin{algorithm}[h]
	\caption{$\mathcal{OP}1:$ Edge deletion}\label{algo_initialguess1}
	\begin{algorithmic}[1]
		\State \textbf{Input:}\,$\{s,t\}=\textbf{indexmax}\left( \tilde{r}^{d,(n)} \right) $, $\Gamma_I^{(n)}$, $\mathcal{D}^{(n)} $  
		\State delete edge $st$, therefore  $\mathcal{D}^{(n)}\leftarrow\mathcal{D}^{(n)}\bigcup \left\{st\right\}$ 
		\If {$d_{s}^{(n)} \neq 1 \lor d_{t}^{(n)} \neq 1 $}
		\State $r^{(n+1)}{(st)} \leftarrow \infty$
		\State Compute $\tilde{r}^d_{s,t}\left(\mathcal{OP}1\right)$ corresponding to $\mathcal{OP}1$
		\State Add the edge back, $\therefore$ $\mathcal{D}^{(n)}\leftarrow\mathcal{D}^{(n)} \setminus \left\{st\right\}$.
		\If{$\left|\tilde{r}^d_{s,t}\left(\mathcal{OP}1\right)\right|<\left|\tilde{r}^{d,(n)}_{s,t}\right|$}
		\State Go to Algorithm-\ref{algo_initialguess2}
		\Else 
		\State Operation $\mathcal{OP}1$ is an \emph{invalid operation}. 
		\State Add the edge back, $\therefore$ $\mathcal{D}^{(n)}\leftarrow\mathcal{D}^{(n)} \setminus \left\{st\right\}$.
		\State $r^{(n+1)}{(st)} \leftarrow r^{(n)}{(st)}$.
		\State Compute $\tilde{r}^d_{s,t}\left(\mathcal{OP}1\right)$ and go to Algorithm-3.
		\EndIf
		\Else
		\State Add the edge back, $\therefore$ $\mathcal{D}^{(n)}\leftarrow\mathcal{D}^{(n)} \setminus \left\{st\right\}$. 
		\State $\tilde{r}^d_{s,t}\left(\mathcal{OP}1\right) \leftarrow \tilde{r}^{d,(n)}_{s,t}$ and go to Algorithm-3.
		\EndIf \label{endop1}
	\end{algorithmic}
\end{algorithm}
\begin{algorithm}[h]
	\caption{$\mathcal{OP}2:$ Increase edge resistance by $1\Omega$}\label{algo_initialguess2}
	\begin{algorithmic}[1]
		\State \textbf{Input:}\,$\{s,t\}=\textbf{indexmax}\left( \tilde{r}^{d,(n)} \right) $, $\Gamma_I^{(n)}$, $\mathcal{D}^{(n)} $
		\If {$r^{(n)}{(st)}<r_{max}$}
		\State 	$r^{(n+1)}(st) \leftarrow r^{(n)}(st) + 1\Omega$
		\State Compute $\tilde{r}^d_{s,t}\left(\mathcal{OP}2\right)$ corresponding to $\mathcal{OP}2$
		\State{$\{\Gamma^{(n+1)}_I, \mathcal{D}^{(n+1)}, A^{(n+1)} \}=\textbf{OPselect}\left(\mathcal{OP}1, \mathcal{OP}2\right)$}{} 
		\State Find node pair, say $\{s_{1},t_{1}\}$, such that $\{s_1,t_1\}=\textbf{indexmax}\{\tilde{r}^{d,(n+1)}\}$, and $s_1t_1 \notin \mathcal{D}^{(n+1)}$.
		\State Go to step \ref{endopx}.
		\Else
		\State Implement $\mathcal{OP}1$, $\therefore$ 
		\State $\Gamma^{(n+1)}_I \leftarrow \Gamma_I^{(n)} \mbox{with committed operation}\,\mathcal{OP}1$
		\State  $\mathcal
		{D}^{(n+1)} \& A^{(n+1)} \leftarrow \mbox{Update}\,\mathcal
		{D}^{(n)} \& A^{(n)}$.
		\State Find node pair, say $\{s_{1},t_{1}\}$, such that $\{s_1,t_1\}=\textbf{indexmax}\{\tilde{r}^{d,(n+1)}\}$, and $s_1t_1 \notin \mathcal{D}^{(n+1)}$.
		\EndIf \label{endopx}
	\end{algorithmic}
\end{algorithm}
\begin{algorithm}[h]
	\caption{Selecting an operation among $\left(\mathcal{OP}1, \mathcal{OP}2\right)$}\label{algo_initialguess5}
	\begin{algorithmic}[1]
		\State \textbf{Input:} $\tilde{r}^d_{s,t}\left(\mathcal{OP}2\right)$, $\tilde{r}^d_{s,t}\left(\mathcal{OP}1\right), \tilde{r}^{d,(n+1)}_{s,t}$
		\State \textbf{Output:} $\Gamma^{(n+1)}_I$, $\mathcal
		{D}^{(n+1)}$ and $A^{(n+1)}$
		\Function {$\textbf{OPselect}\left(\mathcal{OP}1, \mathcal{OP}2\right)$}{}  
		\If{$\left|\tilde{r}^d_{s,t}\left(\mathcal{OP}2\right)\right|<\left|\tilde{r}^{d,(n+1)}_{s,t}\right| \land \left|\tilde{r}^d_{s,t}\left(\mathcal{OP}1\right)\right|<\left|\tilde{r}^{d,(n+1)}_{s,t}\right|$}
		\State Choose and commit an operation which results in $\min{\left\{|\tilde{r}^d_{s,t}\left(\mathcal{OP}1\right)|, |\tilde{r}^d_{s,t}\left(\mathcal{OP}2\right)|\right\}}$
		\ElsIf{$\left|\tilde{r}^d_{s,t}\left(\mathcal{OP}2\right)\right|<\left|\tilde{r}^{d,(n+1)}_{s,t}\right| \land \left|\tilde{r}^d_{s,t}\left(\mathcal{OP}1\right)\right|>\left|\tilde{r}^{d,(n+1)}_{s,t}\right|$}
		\State Choose operation $\mathcal{OP}2$
		\ElsIf{$\left|\tilde{r}^d_{s,t}\left(\mathcal{OP}2\right)\right|>\left|\tilde{r}^{d,(n+1)}_{s,t}\right| \land \left|\tilde{r}^d_{s,t}\left(\mathcal{OP}1\right)\right|<\left|\tilde{r}^{d,(n+1)}_{s,t}\right|$}
		\State Choose operation $\mathcal{OP}1$
		\Else
		\State Go to step-\ref{step13}
		\EndIf \label{endfun}
		\State $\Gamma^{(n+1)}_I \leftarrow \Gamma_I^{(n)} \mbox{with committed operation}$\label{step13}
		\State  $\mathcal
		{D}^{(n+1)}_I \& A^{(n+1)}_I \leftarrow \mbox{Update}\,\mathcal
		{D}^{(n)} \& A^{(n)}$.
		\State \Return $\Gamma_I^{(n+1)}, \mathcal
		{D}^{(n+1)}, A^{(n+1)}$
		\EndFunction
	\end{algorithmic}
\end{algorithm}
\\\noindent \textbf{Case-2:}\,If $\tilde{r}^{d,(n)}_{s,t}<0$ and $st \notin \mathcal{E}_I^{(n)}$, then choose another node pair $s_1,t_1 \in \mathcal{V_B}, \mbox{and}\, s_1t_1 \in \mathcal{E}_I^{(n)}$, across which next maximum absolute resistance distance error occurs. Then, check the sign of $\tilde{r}^{d,(n)}_{s,t}$ to decide an operation to commit.\\
\textbf{Case-3:}\,Now, if $\tilde{r}^{d,(n)}_{s,t}>0 \land st \notin \mathcal{E}_I^{(n)}$, then add an edge $st$ with edge resistance $r^{(n)}(st)=1\Omega$. Edge addition operation for case-$3$ is called as $\mathcal{OP}3$ and is given as Algorithm \ref{algo_initialguess3} $\left(\mbox{at}\,n^{th}\,\mbox{iteration} \right)$. Let $\tilde{r}^d_{s,t}\left(\mathcal{OP}3\right)$ be the resultant resistance distance after committing an operation $\mathcal{OP}3$, if $|\tilde{r}^d_{s,t}\left(\mathcal{OP}3\right)|>|\tilde{r}^{d,(n)}_{s,t}|$ then $\mathcal{OP}3$ is an invalid operation. In case of invalid operation $\mathcal{OP}3$, find another node pair, say $s_1,t_1$, such that $\{s_1,t_1\}=\textbf{indexmax}\left( \tilde{r}^{d,(n)} \right)$.
\begin{algorithm}[h]
	\caption{$\mathcal{OP}3:$ Edge addition}\label{algo_initialguess3}
	\begin{algorithmic}[1]
		\State \textbf{Input:}\, Index $\left\lbrace s,t \right\rbrace $, $\Gamma_I^{(n)}$, $A^{(n)} $
		\If {$ \tilde{r}^{d,(n)}_{s,t}>0$}
		\If{$st \notin \mathcal{E}_I^{(n)}$}
		\State add edge $st$ with $r^{(n)}(st)=1\Omega$ , $\therefore$ $A^{(n+1)}\leftarrow A^{(n)}\bigcup \left\{st\right\}$ 
		\State Compute $\tilde{r}^d_{s,t}\left(\mathcal{OP}3\right)$ 
		\If{$\left|\tilde{r}^d_{s,t}\left(\mathcal{OP}3\right)\right|<\left|\tilde{r}^{d,(n)}_{s,t}\right|$}
		\State Choose operation $\mathcal{OP}3$
		\State $\Gamma^{(n+1)}_I \leftarrow \Gamma_I^{(n)} \mbox{with committed operation}\,\mathcal{OP}3$
		\State  $\mathcal
		{D}^{(n+1)} \& A^{(n+1)} \leftarrow \mbox{Update}\,\mathcal
		{D}^{(n)} \& A^{(n)}$.
		\State Compute $\tilde{r}^{d,(n+1)}$ and find node pair, say $s_1,t_1$, such that $\{s_1,t_1\}=\textbf{indexmax}\{\tilde{r}^{d,(n+1)}\}$.
		\State Go to Step \ref{endop3}
		\Else 
		\State Operation $\mathcal{OP}3$ is an \emph{invalid operation}. 
		\State Remove added edge, $A^{(n+1)}\leftarrow A^{(n+1)} \setminus \left\{st\right\}$
		\State $\Gamma^{(n+1)}_I \leftarrow \Gamma_I^{(n)}$ 
		\State $\mathcal
		{D}^{(n+1)} \& A^{(n+1)} \leftarrow \mbox{Update}\,\mathcal
		{D}^{(n)} \& A^{(n)}$.
		\State Find another node pair, say $\{s_{1},t_{1}\}$, such that $\{s_1,t_1\}=\textbf{indexmax}\{\tilde{r}^{d,(n)}\}$, and $s_1t_1 \notin \mathcal{D}^{(n)}$.
		\State Go to step \ref{endop3}.
		\EndIf
		\Else 
		\State Go to Algorithm-\ref{algo_initialguess4}
		\EndIf 
		\Else 
		\State Implement $\{\mathcal{OP}1,\mathcal{OP}2\}$
		\EndIf \label{endop3}
	\end{algorithmic}
\end{algorithm}
\\\noindent \textbf{Case-4:} If $\tilde{r}^{d,(n)}_{s,t}>0 \land st \in \mathcal{E}_I^{(n)}$, then reduce the edge resistance $r^{(n)}(st)$ by $1\Omega$. Edge addition operation for case-$4$ is called as $\mathcal{OP}4$ and is given as algorithm \ref{algo_initialguess4}.\\
\begin{algorithm}[h]
	\caption{$\mathcal{OP}4:$ Decreasing edge resistance value by $1\Omega$}\label{algo_initialguess4}
	\begin{algorithmic}[1]
		\If {$st \in \mathcal{E}_I^{(n)} \land r^{(n-1)}(st)\ge2\Omega$}
		\State 	$r^{(n+1)}(st) \leftarrow r^{(n)}(st) -1\Omega$
		\State Compute $\tilde{r}^d_{s,t}\left(\mathcal{OP}4\right)$ corresponding to $\mathcal{OP}4$
		\If{$\left|\tilde{r}^d_{s,t}\left(\mathcal{OP}4\right)\right|<\left|\tilde{r}^{d,(n)}_{s,t}\right|$}
		\State Choose operation $\mathcal{OP}4$
		\State $\Gamma^{(n+1)}_I \leftarrow \Gamma_I^{(n)} \mbox{with committed operation}\,\mathcal{OP}4$
		\State Compute set $\tilde{r}^{d,(n+1)}$ and find node pair, say $s_1,t_1$, such that $\{s_1,t_1\}=\textbf{indexmax}\{\tilde{r}^{d,(n+1)}\}$.
		\State Go to step \ref{endop4}
		\Else
		\State $\mathcal{OP}4$ is an \emph{invalid operation}.  
		\State 	$r^{(n+1)}(st) \leftarrow r^{(n)}(st)$
		\State $\Gamma^{(n+1)}_I \leftarrow \Gamma_I^{(n)}$
		\State Find another node pair, say $\{s_{1},t_{1}\}$, such that $\{s_1,t_1\}=\textbf{indexmax}\{\tilde{r}^{d,(n)}\}$, and $s_1t_1 \notin \mathcal{D}^{(n)}$.
		\State Go to step \ref{endop4}.
		\EndIf
		\Else 
		\State Find another node pair, say $\{s_{1},t_{1}\}$, such that $\{s_1,t_1\}=\textbf{indexmax}\{\tilde{r}^{d,(n)}\}$, and $s_1t_1 \notin \mathcal{D}^{(n)}$.
		\EndIf \label{endop4}
	\end{algorithmic}
\end{algorithm}
If both $\mathcal{OP}3$ and $\mathcal{OP}4$ are invalid operations, find another pair $s_1,t_1 \in \mathcal{V_B}$ across which next minimum absolute resistance distance error occurs.\\ 
\indent For every computed node pair across which maximum absolute resistance distance error occurs the sign of corresponding resistance distance error is checked and accordingly operations $\left\{\mathcal{OP}1,\mathcal{OP}2\right\}$ or $\left\{\mathcal{OP}3,\mathcal{OP}4\right\}$ is carried out. \emph{Iterations are carried out till the resistance distance errors in set $\tilde{r}^{d,(n)} \le \epsilon$}, i.e. eventually the resistance distance errors in set $\tilde{r}^{d,(n)}$ does not change significantly. The bound $\epsilon$ can also chosen based on users experience or trial and error. Let the algorithm stops at $n_1$ iteration, then the corresponding network $\Gamma_I^{(n_1)}=\left(\mathcal{G}_I^{(n_1)}, \gamma^{(n_1)}\right)$ is transformed into a $MPRSN$, wherein each edges $ij \in \mathcal{E}_I^{(n_1)}$ is converted to $\mathcal{C}_{ij}$ with appropriate switch positions. Then, the switch positions or the initial guess $\rho^{(0)}$ are extracted from this $MPRSN$.
\section{Modified Auslander, Parter and Goldstein's Algorithm}\label{append:mod_planar}
\begin{figure}[h]
	\centering
	\includegraphics[scale=0.15]{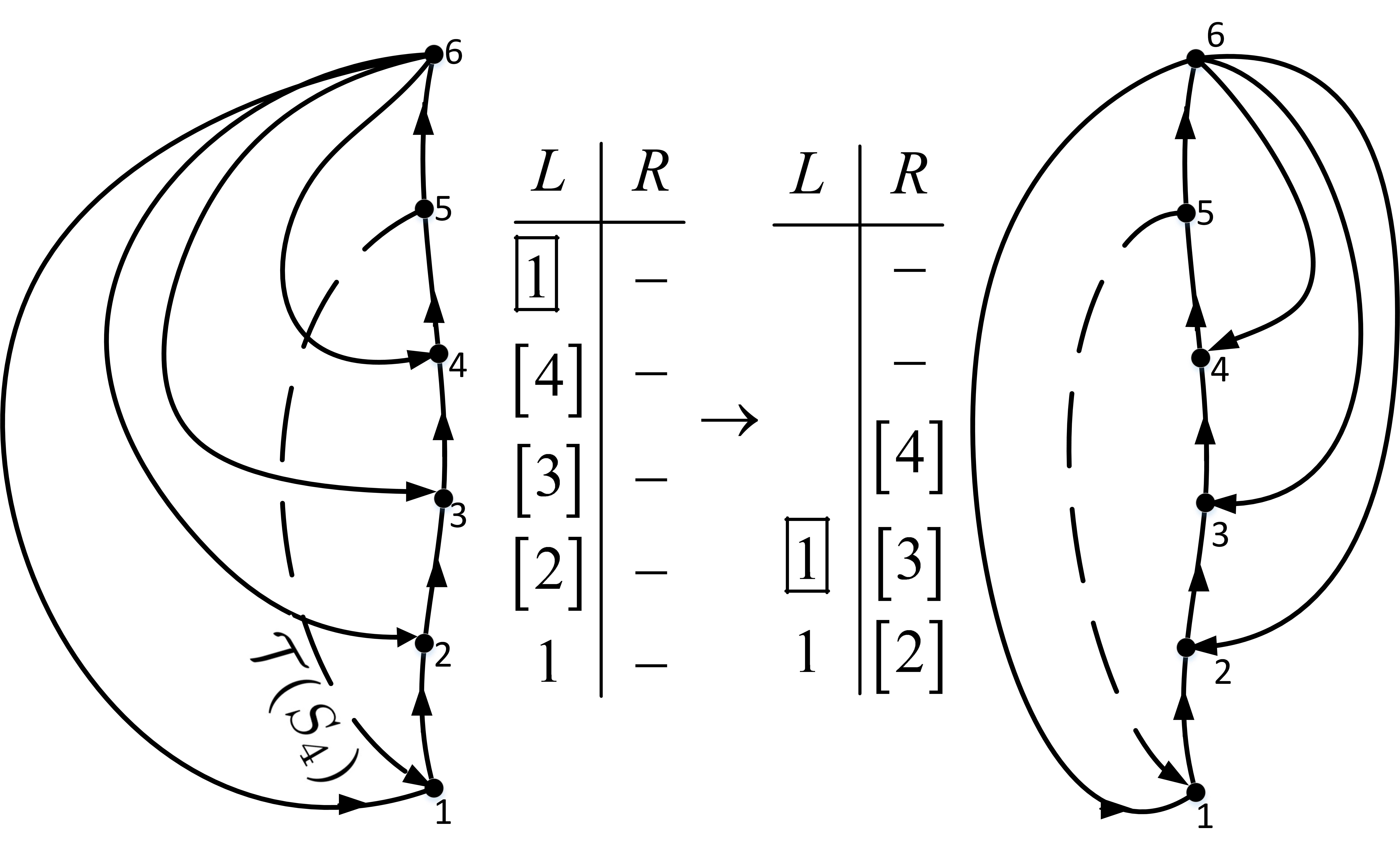}
	\caption{Consider embedding segments $\mathbb{S}_4$ on left side of $\mathcal{T}(c)$, it is seen that Lemma \ref{lemma:three} does not satisfy. Therefore, \textbf{shift some already embedded segments (on left), i.e., $\mathbb{S}_1, \mathbb{S}_2, \mathbb{S}_3$} to the right side of $\mathcal{T}(c)$ to preserve planarity. }
	\label{fig:embed21}
\end{figure}

\noindent We begin by constructing a palm tree representation $P$, where a directed tree $T$ in $P$ is a directed graph with a root vertex such that every vertex in $T$ is reachable from the root. No tree arc enters the root, and exactly one tree arc enters every other vertex. The relation $i \rightarrow^* j$ indicates a path from $i$ to $j$ in $T$. Each vertex in $P$ is associated with two numbers, called \emph{low points}. For the $i^{th}$ vertex, these are denoted as $L_1(i)$ and $L_2(i)$. Additionally, each edge in $P$ is assigned an integer value through a function $\phi: \hat{\mathcal{E}}_{aux} \rightarrow \mathbf{Z}^+$, which orders the adjacency list for efficient processing. For further details on low points and $\phi$, refer to \cite{hopcroft1974efficient, behrensplanarity}. A cycle $c$ in $P$ consists of a sequence of tree arcs and one back edge. Segments not part of $c$ can either be a back edge $i \dashedrightarrow j$ or a path $i \rightarrow^* j$ adjoined by all back edges emanating from $j$. Let $\mathbb{S}$ denote the set of all segments. A path within a segment $\mathbb{S}_k \in \mathbb{S}$ is either a single back edge $i_1 \dashedrightarrow j_1$ or a path $i_1 \rightarrow^* j_1$ with a back edge emanating from $j_1$. To identify a cycle $c$ and paths in each segment, we use a path-finding algorithm based on depth-first search (DFS) and the ordered adjacency list. To accelerate the process, the adjacency list is arranged in increasing order of $\phi(ij), \forall i,j \in \hat{\mathcal{E}}_{aux}$ \cite{behrensplanarity}.
\begin{figure}[h]
	\centering
	\includegraphics[scale=0.2]{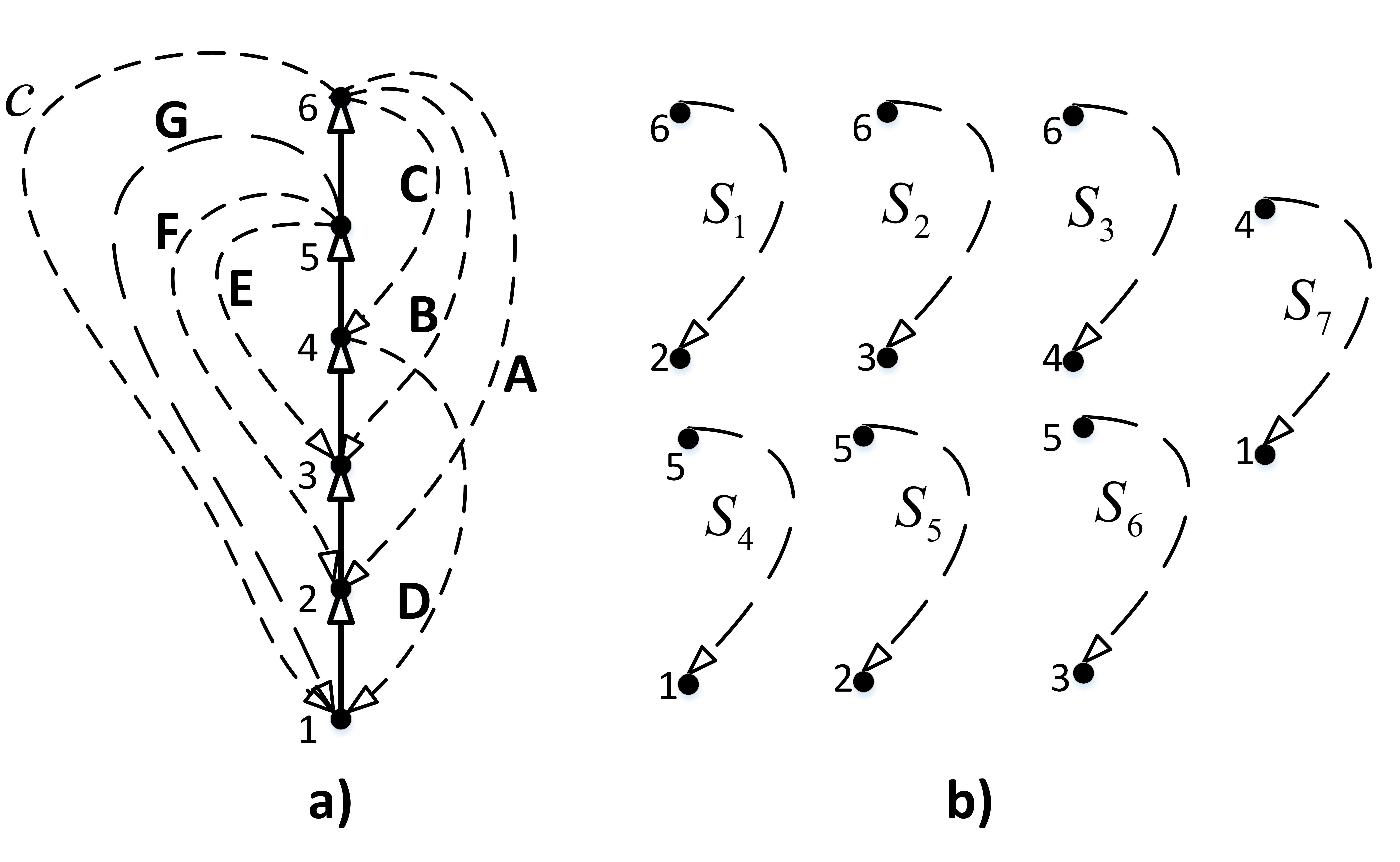}
	\caption{\textbf{a}) Paths generated by path finding algorithm from $P$ in Fig. \ref{fig:palm}  are $c$: 1-2-3-4-5-6-1 A: 6-2 B: 6-3 C: 6-4 D: 4-1 E: 5-3 F: 5-2 G: 5-1. \textbf{b}) segments $\mathbb{S}_1$ to $\mathbb{S}_7$  are obtained after deleting initial cycle $c$ from $P$. }
	\label{fig:palm_compo}
\end{figure}
The algorithm identifies edges using an ordered adjacency list and adds them to the current path. If a back edge is encountered during exploration, it is included in the path, and the search is completed. A detailed explanation of the path-finding algorithm is provided in \cite{hopcroft1974efficient, behrensplanarity}, with an example illustrated in Fig. \ref{fig:palm_compo}(a). The algorithm constructs a cycle $c$, which is then removed from the palm tree $P$, leaving disconnected segments, as shown in Fig. \ref{fig:palm_compo}(b). The path-finding algorithm is also applied to list paths within these segments. Once structuring and path finding are complete, this information is used for planarity testing. Below, we briefly discuss the planarity testing algorithm.\\
\indent The planarity testing algorithm in general does the following,
\begin{itemize}
	\item embed the cycle $c$ on a plane to get $\mathcal{T}(c)$,
	\item embed each segment $\mathbb{S}_k \in \mathbb{S}$ i.e. $\mathcal{T}(\mathbb{S}_k)$ one by one on $\mathcal{T}(c)$. In embedding $\mathbb{S}_k$, which is other than a back edge, we apply path finding algorithm on $\mathbb{S}_k$ and generate all paths. Then, \emph{embed each path one after another on $\mathcal{T}(c)$}. 
	\item Every $\mathcal{T}(\mathbb{S}_k)$ \emph{must go either on the left or right side of $\mathcal{T}(c)$}. When a segment is added to $\mathcal{T}(c)$, certain segments, if needed, are moved from left to right or vice versa to avoid curve crossings. If all $\mathcal{T}(\mathbb{S}_k)$ can be added to $\mathcal{T}(c)$ without any curve crossing, then $\hat{\mathcal{G}}_{aux}$ is said to be planar.
\end{itemize}
For more detailed exposition on planarity testing refer to \cite{hopcroft1974efficient}, for a concise explanation refer to \cite{behrensplanarity}.\\
\indent Further, we present a modification to the Hopcroft, Tarjan, and Goldstein algorithm to extract planar graphs from a non-planar graph. Assume that the cycle $c$ and some segments have already been embedded on the plane, and let $\mathbb{S}_k$ be the next segment to embed. Consider a path $p$ in $\mathbb{S}_k$ to be embedded on $\mathcal{T}(c)$. The following lemma provides a necessary and sufficient condition for embedding $p$:
\begin{lem}\cite{hopcroft1974efficient}\label{lemma:three}
An embedding of a path from $i_1$ to $j_1$ can be added to $\mathcal{T}(c)$ by placing it on left (right) of $\mathcal{T}(c)$ iff \textbf{no} back edge $l \dashedrightarrow k$ that has already been embedded on left satisfies $j_1<k<i_1$. 
\end{lem} 
We use Lemma \ref{lemma:three} to determine the planarity of a graph. To embed a path $\mathcal{T}(p)$, first choose a side (left or right) of $\mathcal{T}(c)$. Suppose $\mathcal{T}(p)$ is placed on the left side of $\mathcal{T}(c)$. Verify if $\mathcal{T}(p)$ satisfies Lemma \ref{lemma:three}. If it does not, this indicates a crossing, and some already embedded segments on the left must be moved to the right side of $\mathcal{T}(c)$ to resolve the conflict. After rearranging, check again whether $\mathcal{T}(p)$ satisfies Lemma \ref{lemma:three}. If satisfied, embed $\mathcal{T}(p)$ on the left side. This process is illustrated in Figure \ref{fig:embed21}.

\begin{figure*}[t]
	\centering
	\subfloat[Place the segment $\mathbb{S}_7$ on both left and right side of $\mathcal{T}(c)$ and find the blocking segments.]{
		\includegraphics[scale=0.15]{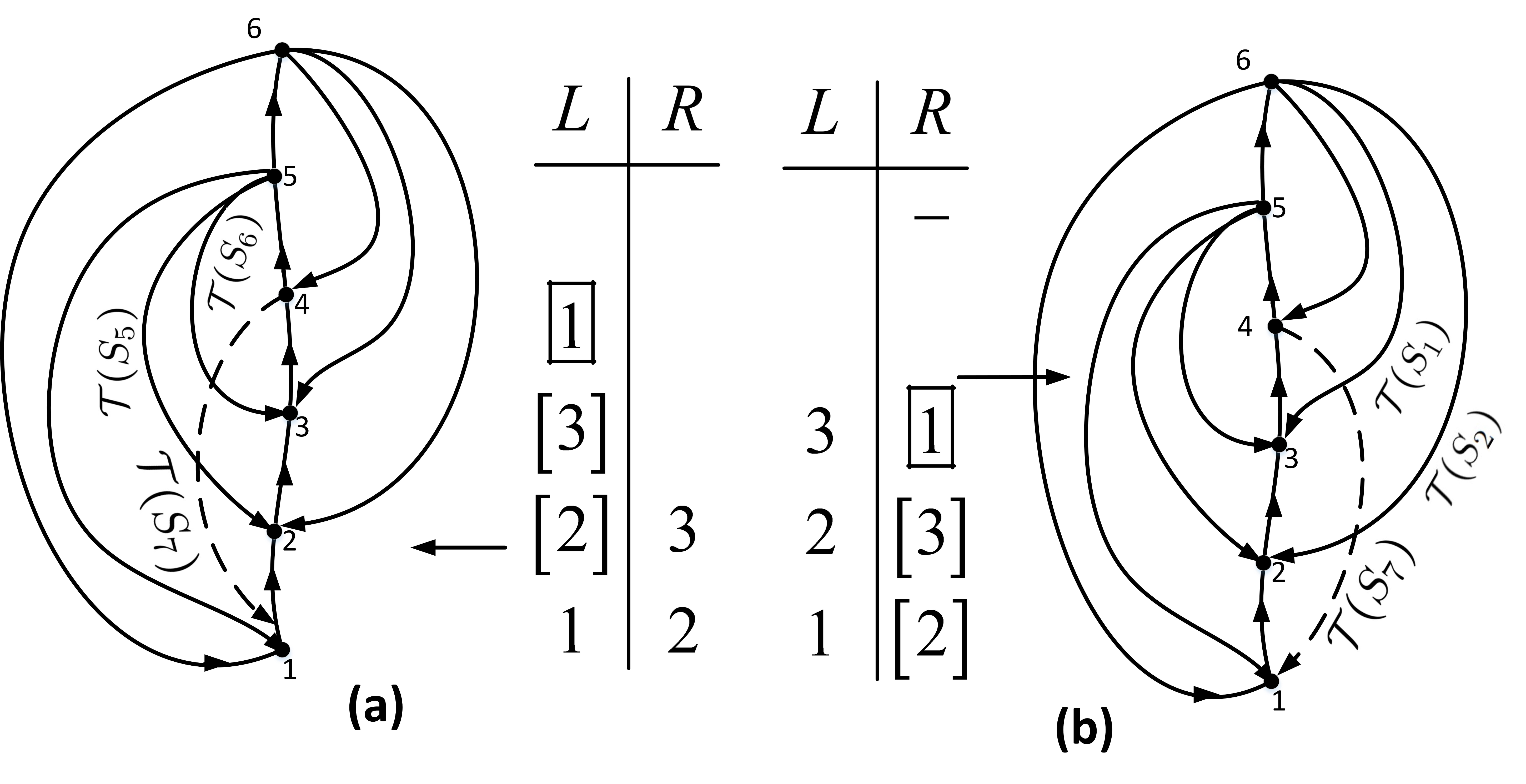}
		\label{fig:embed6}
	}
	\hfil
	\subfloat[Embedding of $\mathcal{G}$, $\mathcal{T(G)}$]{
		\includegraphics[scale=0.15]{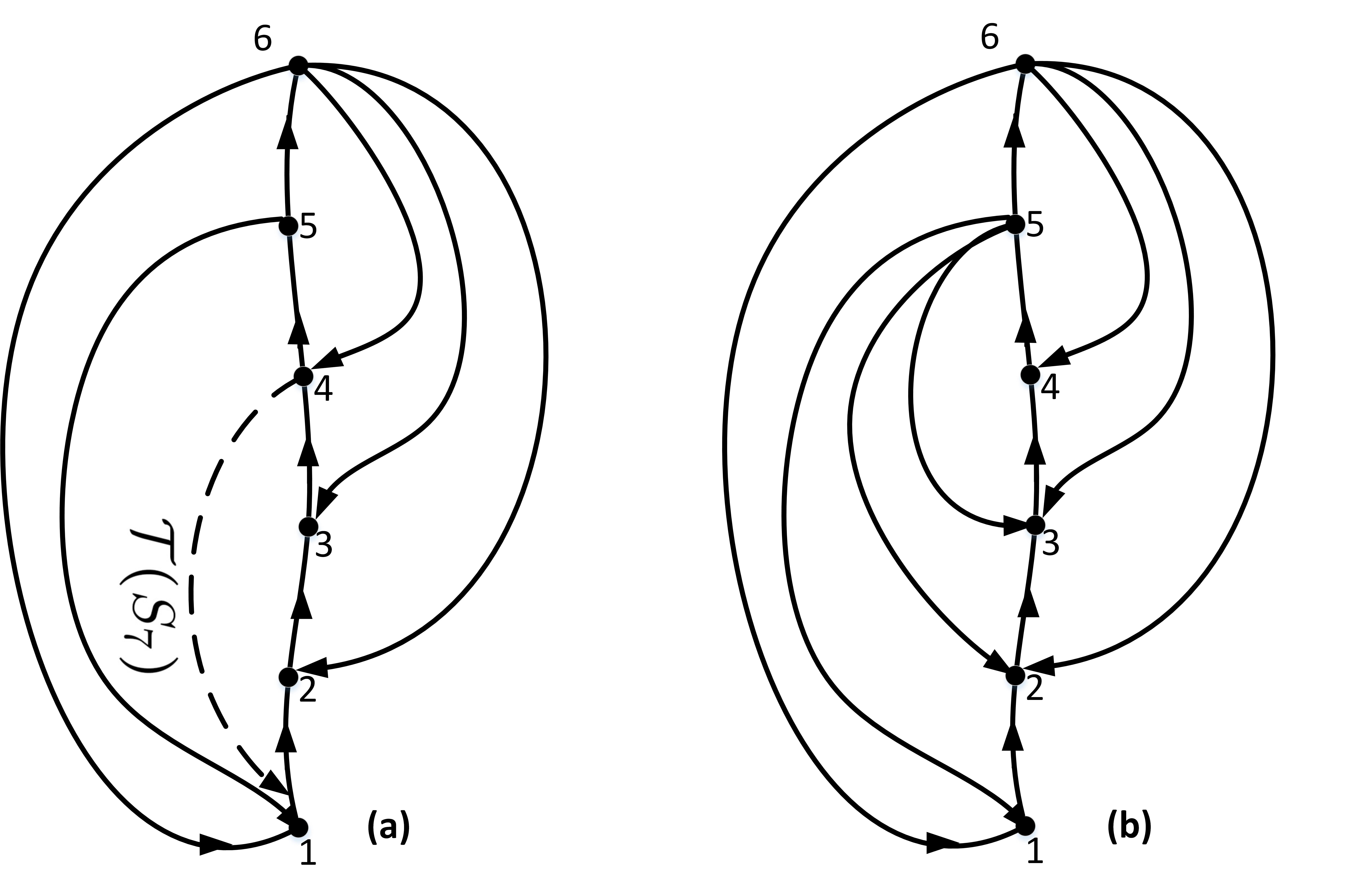}
		\label{fig:embed7}
	}
	\hfil
	\subfloat[Embedding of $\mathcal{G}$, $\mathcal{T(G)}$]{
		\includegraphics[scale=0.15]{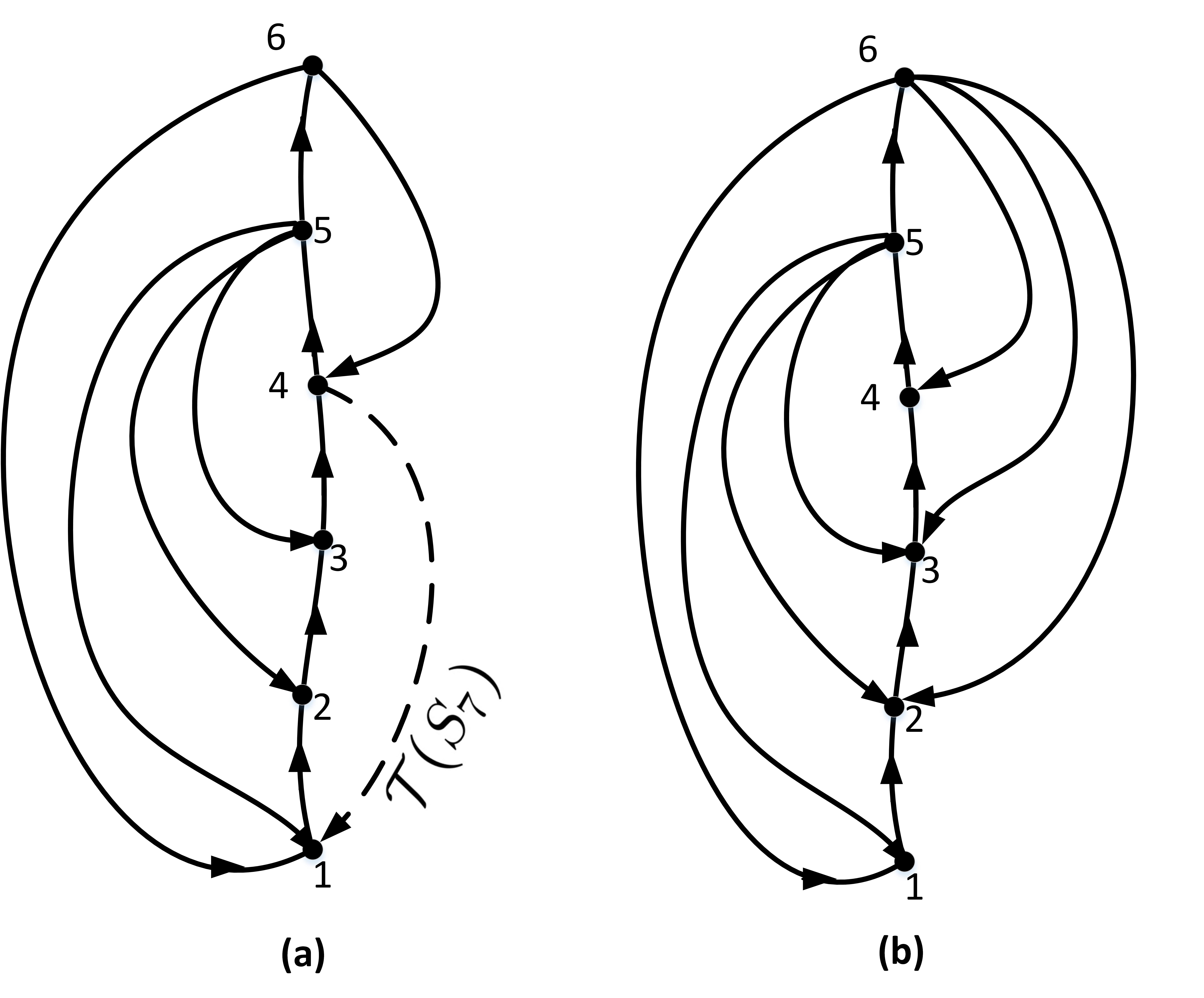}
		\label{fig:embed8}
	}
	\caption{Place the segment $\mathbb{S}_7$ on both left and right side of $\mathcal{T}(c)$ and find the blocking segments. Embeddings of $\mathcal{G}$, $\mathcal{T(G)}$.}
	\label{fig:embeds}
\end{figure*}

\noindent Similarly, we embed each path of $\mathbb{S}_k$ one by one to completely place $\mathcal{T}(\mathbb{S}_k)$ on one side of $\mathcal{T}(c)$. We then move on to embedding the next segment.\\ 
\indent To implement the placement of paths, Hopcroft and Tarjan\cite{hopcroft1974efficient} proposed the usage of data structure stack $L$ and $R$ to save the position of paths and segments during execution. The stack 
$L$ stores  all the vertices $i_k$ such that $1 \rightarrow^* i_k \rightarrow^* i_1$, $1 < i_k < i_1$ and some embedded back edge enters $i_k$ from left. Stack $R$ is defined similarly wherein back edges enter $i_k$ from the right. Implementation of stack $L$ and $R$ is shown in an example in Figure \ref{fig:embed21}. Consider a case of embedding a path, say $\Bar{p}$, of some segment on the left side of $\mathcal{T}(c)$. Update the stacks $L$ and $R$ appropriately and check that the embedding satisfies Lemma \ref{lemma:three}. If it does not satisfy, it means that the embeddings of segments are crossing on the left side. Therefore, shift appropriate segment from left to the right of $\mathcal{T}(c)$ to avoid crossings on left side, and update the entries in $L$ and $R$. Again check whether Lemma \ref{lemma:three} is satisfied on the right side of $\mathcal{T}(c)$. If it does not satisfy then we say that the graph $\hat{\mathcal{G}}_{aux}$ is non planar. At this stage, we extract planar graphs from a non planar graph. The embedding $\mathcal{T}(\Bar{p})$ cannot be placed either on the left or the right side of $\mathcal{T}(c)$. Therefore, 
\vspace{-\topsep}
\begin{itemize}
\item first place $\mathcal{T}(\Bar{p})$ on the left side of $\mathcal{T}(c)$ and update stack $L$.
	\item Find all the instances where the already embedded back edges violate lemma \ref{lemma:three}. We call such back edges as blocking segments and remaining segments as non blocking segments.
	\item The $\mathcal{T}(\Bar{p})$ and blocking segments cannot stay on the left side of $\mathcal{T}(c)$ for maintaining planarity. We therefore construct two planar embeddings, one comprising of $\mathcal{T}(\Bar{p})$ and all non blocking segments and, other containing only blocking segments and non blocking segments.
	\item Then, check whether all the edges of $\mathcal{E}_{aux}$ are present in the constructed planar embeddings. If not, reject that planar  embedding from further analysis.
\end{itemize}	

The above procedure can be understood from an example in 
Figure \ref{fig:embed6}, wherein embedding $\mathcal{T}(\mathbb{S}_7)$ 
leads to non planarity. On the left side, the blocking segments are 
$\left\{\mathcal{T}(\mathbb{S}_5),\mathcal{T}(\mathbb{S}_6)\right\}$ 
and all remaining segments are called the non blocking segments with 
respect to $\mathcal{T}(\mathbb{S}_7)$. Both planar embeddings are 
shown in Fig.\ref{fig:embed7}.
Now, for each planar embeddings check whether all the edges of 
$\mathcal{E}_{aux}$ are present in it. If not, reject that planar 
graph from further analysis. In Fig.\ref{fig:embed7} the edge 
$41 \in \mathcal{E}_{aux}$ is not present, therefore it is not 
considered for further analysis.
Similarly place $\mathcal{T}(\mathbb{S}_7)$ on the right 
side of $\mathcal{T}(c)$ as shown in Fig.\ref{fig:embed6}. 
Find the blocking and non blocking edges with respect to the 
$\mathcal{T}(\mathbb{S}_7)$. The segments $\{\mathcal{T}(\mathbb{S}_2), 
\mathcal{T}(\mathbb{S}_1)\}$ are the blocking segments. The planar 
embeddings corresponding to this is shown in Fig.\ref{fig:embed8}
\noindent A generalised algorithm is given in Algorithm \ref{alg:planar_nplanar}. Every time a path is to be embedded, Algorithm \ref{alg:planar_nplanar} is invoked.
\begin{algorithm}
	\caption{Constructing planar embeddings from a non embedding}\label{alg:planar_nplanar}
	\begin{algorithmic}[1]
		\State \textbf{Input:} $c$, $\mathbb{S}_j$.
		\State \textbf{Output:} Set of admissible planar embeddings.
		\Require $L^{(1)}$ \& $R^{(1)}$ are empty stack,
		\State Compute all paths in segment $\mathbb{S}_j$
		\For {$i \le \mbox{total number of paths in}\,\mathbb{S}_j$}
		\State {place $\mathcal{T}(p_i)$ on $\mathcal{T}(c)$ along with already embedded segments. Where $p_i$ is $i^{th}$ path in $\mathbb{S}_j$.}
		\If {\texttt{embedding is planar}}
		\State {$L^{(i)}\leftarrow \mbox{Update}\,L^{(i-1)}$ and $R^{(i)}\leftarrow\mbox{Update}\,R^{(i-1)}$} 
		\Else 
		\State Place $\mathcal{T}(p_i)$ on left side of $\mathcal{T}(c)$
		\State {Update $L^{(i-1)}$ and $R^{(i-1)}$ to construct $L^{(i)}$ and $R^{(i)}$} \vspace{0.1cm}
		\State Find blocking segments \vspace{0.1cm}
		\State {Construct planar embeddings wherein embedding of blocking segments and $\mathcal{T}(p_i)$ are not together.}\vspace{0.1cm}
		\State {Check whether all the edges in $\mathcal{E}_{aux}$ are contained in the constructed planar embeddings. If not, reject such planar embeddings from further analysis.}\vspace{0.1cm}
		\State {Now, delete $\mathcal{T}(p_i)$ on the left side and place $\mathcal{T}(p_i)$ on right side of $\mathcal{T}(c)$}\vspace{0.1cm} 
		\State {Update $L^{(i-1)}$ and $R^{(i-1)}$ to construct $L^{(i)}$ and $R^{(i)}$}\vspace{0.1cm} 
		\State Find blocking segments\vspace{0.1cm}
		\State {Construct planar embeddings wherein embedding of blocking segments and $\mathcal{T}(p_i)$ are not together}\vspace{0.1cm}
		\State {Check whether all the edges in $\mathcal{E}_{aux}$ are contained in the constructed planar embeddings. If not, reject such planar embeddings from further analysis.}\vspace{0.1cm}
		\EndIf
		\State $i \leftarrow i+1$
		\EndFor
		\State Construct a set of admissible planar embeddings. 
	\end{algorithmic}
\end{algorithm}
\end{document}